\RequirePackage{fix-cm}
\documentclass[twocolumn]{svjour3}          
\smartqed  
\usepackage{mathptmx}      
\usepackage{graphicx,color} 
\usepackage{amsmath}
\usepackage{epstopdf}
\usepackage{textgreek}
\usepackage{amssymb}
\usepackage{caption}

\newcommand{\um}{\textmu m\,}

\newcommand{\minute}{\ensuremath{^\prime}\,}
\newcommand{\second}{\ensuremath{^{\prime\prime}}\,}

\newcommand{\Planck}{{\it Planck}}  %
\newcommand{\Herschel}{{\it Herschel}}  %
\newcommand{\Pacs}{{\it PACS}}  %
\newcommand{\Estadius}{{\it Estadius}}  %
\newcommand{\BICEPtwo}{{\it BICEP2}}  %
\newcommand{\Pilot}{{\it PILOT}}  %
\newcommand{\PILOT}{{\it PILOT}}  %
\newcommand{\ICS}{ICS}  %
\newcommand{\HWP}{HWP}  %

\newcommand{\taucal}{\tau_{cal}}  
\newcommand{\taubolo}{\tau_{det}} 
\newcommand{\taugli}{\tau_{gli}}   
\newcommand{\tauics}{\tau_{ics}}  
\newcommand{\RespICS}{\rho_{ICS}}  
\newcommand{\DeltaICS}{\Delta^{ICS}_{\rm on-off}}  
\newcommand{\IICS}{I_{\rm ICS}}  
\newcommand{\IICSON}{I_{\rm ON}}  
\newcommand{\IICSOFF}{I_{\rm OFF}}  
\newcommand{\RICS}{R_{\rm ICS}}  
\newcommand{\RICSref}{R_{\rm ref}}  
\newcommand{\IICSref}{I_{\rm ref}}  
\newcommand{\VICS}{V_{\rm ICS}}  
\newcommand{\TFP}{T_{\rm 300}}  
\newcommand{\TMirror}{T_{\rm M1}}  
\newcommand{\Tfen}{T_{\rm win}}  

\newcommand{\hwppos}{HWP_{\rm pos}}  

\newcommand{\mic}{\,{\rm \mu m} }
\newcommand{\degr}{{\rm ^o} }

\newcommand{\TRANS}{TRANS}  
\newcommand{\REFLEX}{REFLEX}  

\newcommand{\flightone}{flight\#1}
\newcommand{\flighttwo}{flight\#2}

\newcommand{\SNRp}{SNR_p}

\newcommand{\StokesI}{I}                    
\newcommand{\StokesQ}{Q}                    
\newcommand{\StokesU}{U}                    
\newcommand{\polfrac}{p}                    
\newcommand{\polang}{\psi}                  

\begin{document}

\title{Inflight Performance of the PILOT balloon-borne experiment
}

\titlerunning{PILOT inflight performance}        

\author{
A. Mangilli,
G. Fo\"enard,
J. Aumont,
A. Hughes,
B. Mot,
J-Ph. Bernard,
A. Lacourt,
I. Ristorcelli,
L. Montier,
Y. Longval,
P. Ade,
Y. Andr\'e,
L. Bautista,
P. deBernardis,
O. Boulade,
F. Bousqet,
M. Bouzit,
N. Bray,
V. Buttice,
M. Charra,
M. Chaigneau,
B. Crane,
E. Doumayrou,
J.P. Dubois,
X. Dupac,
C. Engel,
P. Etcheto,
Ph. Gelot,
M. Griffin,
S. Grabarnik,
P. Hargrave,
Y. Lepennec,
R. Laureijs,
B. Leriche,
S. Maestre,
B. Maffei,
J. Martignac,
C. Marty,
W. Marty,
S. Masi,
F. Mirc,
R. Misawa,
J.M. Nicot,
J. Montel,
J. Narbonne,
F. Pajot,
E. P\'erot,
G. Parot,
J. Pimentao,
G. Pisano,
N. Ponthieu,
L. Rodriguez,
G. Roudil,
H. Roussel,
M. Salatino,
G. Savini,
O. Simonella,
M. Saccoccio,
S. Stever,
P. Tapie,
J. Tauber,
C. Tibbs,
C. Tucker
}

\authorrunning{A. Mangilli et al.} 

\institute{
A. Mangilli, G. Fo\"enard, J. Aumont, A. Hughes, J-Ph. Bernard, A. Lacourt, C. Marty, W. Marty, R. Misawa ,
B. Mot, L. Montier, J. Narbonne, F. Pajot, I. Ristorcelli, G. Roudil, L. Bautista
 \at
Institut de Recherche en Astrophysique et Planetologie
(IRAP), Universit\'e Paul Sabatier,
9 Av du Colonel Roche, BP 4346, 31028 Toulouse cedex 4; \\
\email{Anna.Mangilli@irap.omp.eu}           
\and
C. Engel
\at
LAM
38 rue F. Joliot-Curie
13388 Marseille CEDEX 13, France; \\ 
\and
M. Bouzit, V. Buttice, M. Charra, B. Crane, J.P. Dubois, B. Leriche,
Y. Longval, B. Maffei, S. Stever, M. Chaigneau
\at
Institut d'Astrophysique Spatiale (IAS), B\^at 121, Universit\'e
Paris XI, Orsay, France; \\
\and
P. Ade, M. Griffin, P. Hargrave, G. Pisano, G. Savini, C. Tucker
\at
Department of Physics and Astrophysics, PO BOX 913, Cardiff University, 
5 the Parade, Cardiff, UK; \\
\and
Y. Andr\'e, F. Bousqet, S. Maestre, J. Montel,O. Simonella,
M. Saccoccio, P. Gelot, J.M. Nicot, P. Tapie,F. Mirc,N. Bray,
P. Etcheto, G. Parot
\at
Centre National des Etudes Spatiales, DCT/BL/NB, 18 Av. E.
Belin, 31401  Toulouse, France; \\
\and
P. deBernardis, S. Masi, M. Salatino, J. Pimentao
\at
Universita degli studi di Roma "La Sapienza", Dipartimento di Fisica, 
P.le A. Moro, 2, 00185, Roma, Italia; \\
\and
O. Boulade, E. Doumayrou, L. Rodriguez, Y. Lepennec, J. Martignac
\at
CEA/Saclay, 91191 Gif-sur-Yvette Cedex, France; \\
\and
S. Grabarnik, R. Laureijs, J. Tauber, C. Tibbs, X. Dupac
\at
Scientific Support Office, SRE-S, ESTEC, PO Box 299,2200AG Noordwijk, The Netherlands;\\
\and
N. Ponthieu
\at
Grenoble University, Grenoble, France; \\
\and
E. P\'erot
\at
Thales Services, Toulouse, France; \\
\and
G. Savini
\at 
Department of Physics \& Astronomy, University College London, Gower
Place, London WC1E 6BT, United\,Kingdom;\\
\and
H. Roussel
\at 
Institut d'Astrophysique de Paris
98 bis boulevard Arago, 75014 Paris, France;\\
}

\date{Received: date / Accepted: date}

\maketitle

\begin{abstract}

The Polarized Instrument for Long-wavelength Observation of the Tenuous interstellar medium ({\Pilot}) is a balloon-borne experiment that aims to measure the polarized emission of thermal dust at a wavelength of 240\,\um (1.2 THz). A first {\Pilot} flight of the experiment took place from Timmins, Ontario, Canada, in September 2015 and a second flight took place from Alice Springs, Australia in April 2017. In this paper, we present the inflight performance of the instrument.  Here we concentrate on the instrument performance as measured during the second flight, but refer to the performance observed during the first flight, if it was significantly different. We present a short description of the instrument and the flights. We measure the time constants of the detectors using the decay of the observed signal during flight following high energy particle impacts (glitches) and switching off the instrument's internal calibration source. We use these time constants to deconvolve the timelines and analyze the optical quality of the instrument as measured on planets. We then analyze the structure and polarization of the instrumental background. We measure the detector response flat field and its time variations using the signal from the residual atmosphere and from the internal calibration source. Finally, we analyze the spectral and temporal properties of the detector noise.  The inflight performance is found to be satisfactory and globally in line with expectations from ground calibrations. We conclude by assessing the expected inflight sensitivity of the instrument in light of the measured inflight performance.

\end{abstract}

\section{Introduction}
\label{sec:introduction}

Interstellar dust grains account for $\simeq$1\% of the mass of the interstellar medium (ISM).  They are involved in different important processes such as photo-electric heating of the neutral interstellar gas, cooling in dense star-forming regions and the formation of molecules, including H$_2$ at their surface.  Dust emission can be used to trace the structure of the interstellar medium (ISM) in the Milky Way and in the local Universe (e.g., \cite{foyleetal12,combesetal12,hilletal12}). The thermal dust emission can be modeled using a modified blackbody spectrum in the infrared to submillimeter wavelength range. ISM dust grains absorb starlight in the visible and ultra-violet, which heats them to temperatures of $\simeq$17\,K in the diffuse ISM of our Galaxy. The polarization of thermal dust emission is believed to arise from the irregular shape of dust grains. As the grains rotate, they align their minor axis to the local magnetic field (e.g., \cite{Lazarian2003,Lazarian2007}). This partial alignment causes a fraction of their thermal emission to be linearly polarized in a direction orthogonal to the magnetic field direction as projected on the sky. For the same reason, non-polarized starlight passing through aligned dust grains also becomes polarized: preferential absorption along the long axis of the grains leads to extinction that is polarized parallel to the magnetic field lines.

The balloon experiment Archeops (\cite{Benoit2004a}) mapped the polarized dust emission at 353\,GHz with $\sim13^\prime$ resolution over $\sim20$\% of the sky.  These measurements indicated high polarization levels (up to 15\%) in the diffuse ISM.  More recently, the Planck satellite has mapped the polarized dust emission in the wavelength range from $850\mic$ (353\,GHz) to 1.0\,cm (30\,GHz) over the entire sky (\cite{planck2014-XIX}).  The {\Planck} satellite data confirmed the existence of highly polarized regions at high galactic latitudes with polarization fractions up to 20\%. As a consequence, polarized dust thermal emission is a dominant foreground contaminant to the observation of the Cosmic Microwave Background (CMB) polarization (see, \cite{BICEP2Bmodes}). The goal of the {\PILOT} observations is to improve our understanding of the thermal dust polarization signal, by measuring it at higher frequencies in the far-infrared.

We briefly describe the {\PILOT} instrument in Sect.\,\ref{sec:pilot_instrument} and the observations during flight in Sect.\,\ref{sec:flights}.
The use of glitches and of the instrument internal calibration source to measure detector time constants is described in Sect.\,\ref{sec:glitches}. The inflight performance of the instrument's optics is discussed in Sect.\,\ref{sec:psf}. The instrumental background observed during flight as well as its apparent polarization is discussed in Sect.\,\ref{Sec:Inst_bkd}. The detector response measured on the residual atmospheric signal and detector noise properties are given in Sect.\,\ref{Sec:resp} and Sect.\,\ref{Sec:noise} respectively. Section\,\ref{Sec:sensitivity} summarizes the expected instrument sensitivity, given the measured inflight performance. In Section\,\ref{sec:conclusion}, we present our conclusions.

\section{The {\Pilot} instrument}
\label{sec:pilot_instrument}

\subsection{Instrument description}
\label{sec:Instrument_description}

A complete description of the {\Pilot} instrument is available in \cite{Bernard_etal2016}.  Here, we give a brief description for completeness. Table \ref{tab:summarize_optics} summarizes the main characteristics of the instrument.\\

The optics of the instrument is composed of an off-axis paraboloid primary mirror (M1) with diameter of 0.83\,m and an off-axis ellipsoid secondary mirror (M2). The combination respects the Mizugushi-Dragone condition to minimize depolarization effects (see \cite{Bernard_etal2016,Engel2013}).  All optics following M1, including M2, is cooled at cryogenic temperature of 2\,K.

Following the Gregorian telescope, the beam is folded using a flat mirror (M3) towards a re-imager and a polarimeter.  Two lenses (L1 and L2) are used to re-image the focus of the telescope on the detectors. A Lyot-stop is placed between the lenses at a pupil plane that is a conjugate of the primary mirror. A rotating Half-Wave Plate ({\HWP}), made of Sapphire, is located next to the Lyot-stop.  The bi-refringent material of the {\HWP} introduces a phase shift between the two orthogonal components of the incident light. A polarization analyzer consisting of parallel metallic wires is placed at a $45\degr$ angle in front of the detectors, in order to transmit one polarization to the transmission ({\TRANS}) focal plane and reflect the other polarization to the reflection ({\REFLEX}) focal plane.  Observations at two (or greater than two) different {\HWP} angles allow us to reconstruct the Stokes parameters I, Q and U, as described in Sect.\,\ref{sec:Polarisation_measurements}.  Each of the {\TRANS} and {\REFLEX} focal planes include 1024 bolometers (4 arrays of 16 X 16 pixels). They are cooled to 300\,mK  by a closed cycle $^3$He fridge. The detectors were developed by CEA/LETI for the {\Pacs} instrument on board the {\Herschel} satellite.

In order to reconstruct the pointing of the instrument, we use the {\Estadius} stellar sensor developed by CNES for stratospheric applications and described in \cite{Montel+2015}.  This system provides an angular resolution of a few arcseconds, which is required to optimally combine observations of the same part of the sky obtained with various polarization analysis angles. A key feature of {\Estadius} is that remains accurate even with fast scan speeds (up to  1\,$\degr$/s).  An internal calibration source ({\ICS}) is used inflight to calibrate time variations of the detector responses. This device is described in \cite{Hargrave2006,Hargrave2003}. The source is located behind mirror M3 and illuminates all detectors simultaneously. It is driven using a square modulated current. The current and voltage of the source are measured permanently during flight, in order to monitor the power dissipated in the source.

\begin{table*}[ht]
\caption{\label{tab:summarize_optics}
Main optical characteristics of the {\PILOT} instrument.}
\begin{center}
\begin{tabular}{|l|c|c|}
\hline\noalign{\smallskip}
\hline
Telescope type & \multicolumn{2}{|c|}{Gregorian} \\
Equivalent focal length [mm] & \multicolumn{2}{|c|}{1790} \\
Numerical aperture & \multicolumn{2}{|c|}{$F/2.5$} \\
FOV [$\degr$] & \multicolumn{2}{|c|}{$1.0 \times 0.8$} \\
Ceiling altitude & \multicolumn{2}{|c|}{$\sim$3 hPa} \\
Pointing reconstruction & \multicolumn{2}{|c|}{translation$=1{\second}$, rotation$=6{\second}$, $1\sigma$}\\
Gondola mass & \multicolumn{2}{|c|}{$\sim$1100 kg} \\
\hline
\hline
Primary mirror type & \multicolumn{2}{|c|}{Off-axis parabolic}\\
Primary mirror dimension [mm] & \multicolumn{2}{|c|}{$930 \times 830$}\\
M1 used surface projected diameter [mm] & \multicolumn{2}{|c|}{730}\\
Focal length [mm] & \multicolumn{2}{|c|}{750} \\
\hline
\hline
Detector type & \multicolumn{2}{|c|}{Multiplexed bolometer arrays} \\
Total number of detectors  & \multicolumn{2}{|c|}{2048} \\
Detectors temperature [mK] & \multicolumn{2}{|c|}{300} \\
Sampling rate [Hz] & \multicolumn{2}{|c|}{40}\\
\hline
\hline
Photometric channels & SW Band & LW Band \\
$\lambda_0$ [$\mic$] & $240$ &  $550$ \\
$\nu_0$ [GHz] & $1250$ &  $545$ \\
$\Delta\nu/\nu$ & $0.27$ & $0.31$ \\
beam FWHM [{\minute}] & 1.9 &  3.3 \\
Minimum Strehl ratio & 0.95 & 0.98 \\
\hline
\noalign{\smallskip}\hline
\end{tabular}
\end{center}
\end{table*}

\subsection{Polarization measurements}
\label{sec:Polarisation_measurements}

Assuming a perfect {\HWP}, the {\Pilot} measurements $m$ are related to the input Stokes parameters $\StokesI$, $\StokesQ$, $\StokesU$ of partially linearly polarized light through
\begin{equation}
m = R_{xy} T_{xy} \times [\StokesI \pm \StokesQ_{inst} \cos4\omega \pm \StokesU_{inst} \sin4\omega]+O_{xy},
\label{eq:pol_measure_easy_Stokes}
\end{equation}
where $R_{xy}$ and $T_{xy}$ are the system response and optical transmission respectively, and $O_{xy}$ is an arbitrary electronics offset.  For the configuration of the {\HWP} and polarizer in the instrument, $\omega$ is the angle between the {\HWP} fast axis direction and the horizontal direction measured counterclockwise as seen from the instrument. The $\pm$ sign is $+$ and $-$ for the {\REFLEX} and {\TRANS} arrays respectively (see \cite{Bernard_etal2016}). Note that, with the above conventions, $\StokesQ_{inst}$ and $\StokesU_{inst}$ are defined with respect to instrument coordinates in the IAU convention, with $\StokesQ_{inst}$=0 for vertical polarization.
{For {\PILOT}, $\omega$ is related to a mechanical {\HWP} position called $\hwppos$, which can be varied continuously over the range 1 $<\hwppos<$ 8 as
\begin{equation}
\omega = 87.25\degr -(\hwppos - 5) \times 11.25\degr,
\label{eq:pol_hwp_angle}
\end{equation}
allowing the {\HWP} fast axis to vary by $\pm 45\degr$ around the vertical direction.}
When referring to the sky polarization $\StokesQ$ and $\StokesU$, Equ.\,\ref{eq:pol_measure_easy_Stokes} becomes
\begin{equation}
m = R_{xy} T_{xy} \times [\StokesI \pm \StokesQ \cos(2\theta) \pm \StokesU \sin(2\theta)]+O_{xy},
\label{eq:pol_measure_easy_Stokes_sky}
\end{equation}
where $\theta=2\times\omega+\phi$ is the analysis angle, $\phi$ is the time varying parallactic angle measured counterclockwise from equatorial North to Zenith for the time and direction of the current observation, and $\StokesQ$ and $\StokesU$ are in the IAU convention with respect to equatorial coordinates. In practice, maps of $\StokesQ$ and $\StokesU$ are derived from observing the same patch of sky with at least two values of the analysis angle taken at different times in general. Inversion to derive sky maps of $\StokesI$, $\StokesQ$ and $\StokesU$ can be done through polarization map-making algorithms (see for instance \cite{deGasperis+2005}). The light polarization fraction $\polfrac$ and polarization direction $\polang$ are then defined as:
\begin{equation}
\label{equ:pem}
\polfrac=\frac{\sqrt{\StokesQ^2+\StokesU^2}}{\StokesI}
\end{equation}
and
\begin{equation}
\label{equ:thetam}
\polang=0.5 \times \arctan(\StokesU/\StokesQ).
\end{equation}

\section{The {\Pilot} flights and observations}
\label{sec:flights}

{\Pilot} is carried to the stratosphere by a generic gondola suspended under an open stratospheric balloon through a flight chain. The flights are operated by the French National Space Agency (CNES). {\Pilot} uses a 803Z class balloon, with a helium gas volume of $\sim$ 800 000 $m^3$ at ceiling altitude. 
 
The instrument can be pointed towards a given sky direction using the gondola rotation around the flight chain and rotation of the instrument around the elevation axis (see \cite{Bernard_etal2016}). Scientific observations are organized into individual observing tiles (also called observations for short) during which a given rectangular region of the sky is scanned by combining the azimuth and elevation rotations.

The flight plan is built taking into account the various observational constraints such as the visibility of astronomical sources, the minimum allowed angular distance between the instrument optical axis and bright sources such as the Sun or the Moon, elevation limits due to the presence of the Earth at low elevations and the balloon at high elevations. The expected performance of the instrument is taken into account when establishing the flight plan, in order to distribute the observing time according to the science objectives, and to evenly distribute both the polarization analysis directions (angle $\theta$ in Eq.\,\ref{eq:pol_measure_easy_Stokes_sky}) and the scanning directions for any given astronomical target.

\subsection{{\flightone}}
\label{sec:flight_one}

The first flight of the {\Pilot} experiment took place from the launch-base facility at the airport of Timmins, Ontario, Canada on September 21, 2015 at 9:00 PM local time. The launch was part of a campaign led by CNES and the Canadian Space Agency (CSA), during which six stratospheric balloon experiments were successfully conducted.

\begin{figure*}[ht]
\begin{center}
\includegraphics[width=0.7\textwidth]{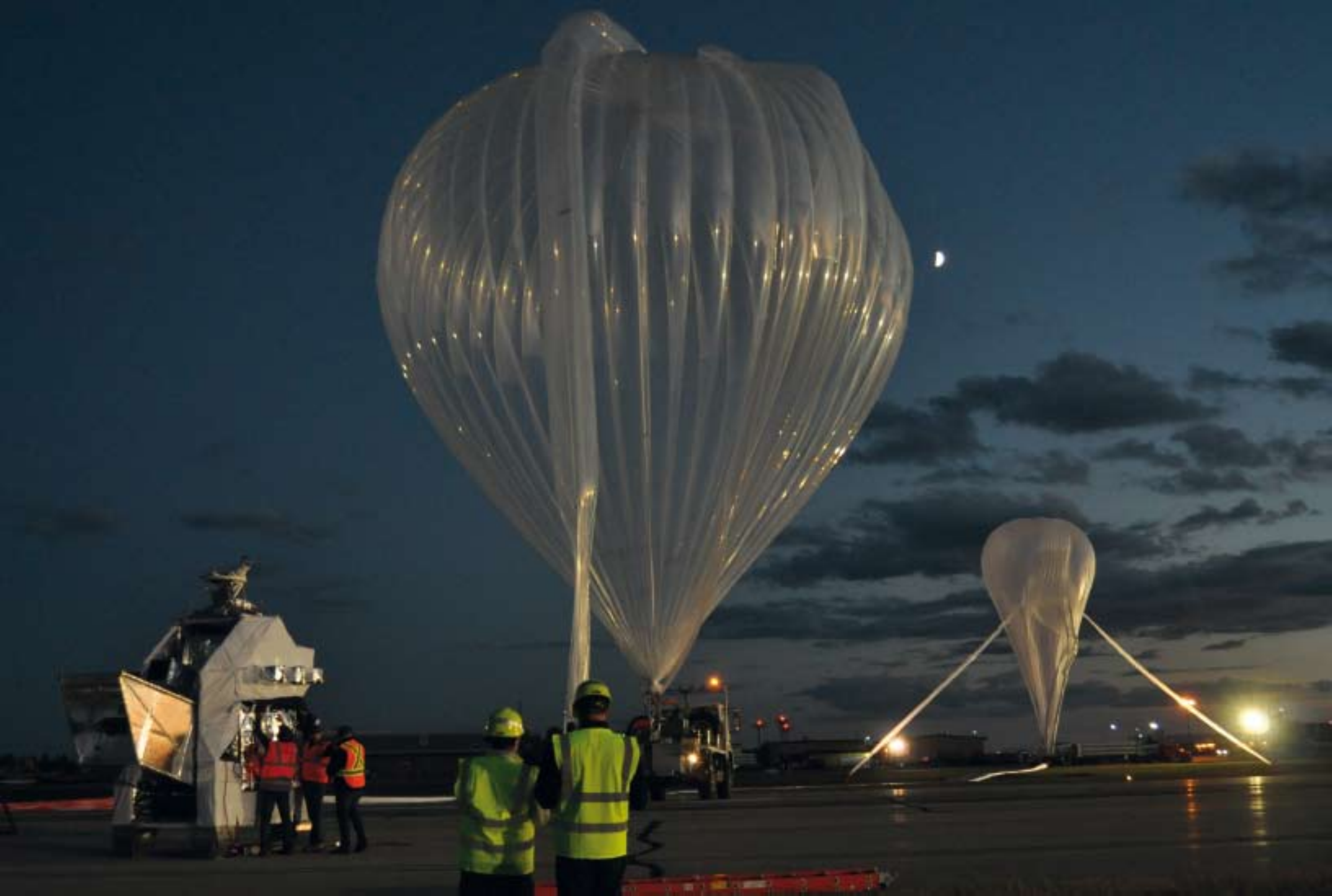}\caption{\label{fig:launch} The {\Pilot} experiment on the tarmac before launch for the first flight from Timmins, Ontario, Canada on September 21 2015.}
\end{center}
\end{figure*}

The ascent to the stratosphere took 2.5\,hr. At ceiling, the altitude of the experiment remained relatively stable between 38\,km and 39\,km over a time period of 18\,hr. The total duration of the flight was 24\,hr allowing us to obtain about 15\,hr of science data. Approximately 4\,hr were spent optimizing the detector readout chain settings, recycling of the $^3$He fridge and slewing between individual observing regions. The focal plane temperature remained stable during the whole ceiling period, at a temperature of $\simeq321\,mK$ and $\simeq325\,mK$ for the {\TRANS} and {\REFLEX} focal planes respectively. Out of the eight bolometer arrays, array \#1 ({\TRANS}) and array \#3 ({\REFLEX}) were not operational during
{\flightone} due to an electrical short circuit affecting the clock signal used for the time-domain multiplexing of the arrays, a problem which was already encountered during ground calibrations (see \cite{Misawa_etal2017}).

The scientific observations performed during {\flightone} are summarized in Tab.\,\ref{tab:flight1-observations}.  During this flight, scientific observations were performed scanning the telescope at constant elevation, in order to minimize residual atmospheric emission.  We collected a total of 5.5\,hr of science data on star-forming regions, 2.4\,hr on cold cores, 1.4\,hr on external galaxies, and 4.6\,hr on a region of the sky with low brightness.  Calibration data were obtained on the planets Uranus and Saturn. We also obtained `skydip' measurements, during which we explored the whole range of allowed
elevations, in order to characterize the residual atmospheric emission (see Sect. \ref{Sec:ICS_resp}).

\begin{table*}[ht]
\centering
\caption{ Summary of the observations obtained during {\flightone}.}
\label{tab:flight1-observations}
\begin{tabular}{llll}
\hline
\multicolumn{1}{c}{Source} &
\multicolumn{1}{c}{\begin{tabular}[c]{@{}c@{}}Observation Time \\
                     {[}mn{]}\end{tabular}} &
\multicolumn{1}{c}{\begin{tabular}[c]{@{}c@{}}Map size\\ {[}deg $\times$
                     deg{]}\end{tabular}} &
\begin{tabular}[c]{@{}l@{}}Total depth\\ {[}$deg^2/h${]}\end{tabular} \\
\hline
Taurus         & 117      & 12 $\times$ 8  & 55 \\ 
Orion           & 145.3  & 10 $\times$ 10 & 47.8 \\
Aquila Rift   & 46       & 8 $\times$ 8      & 94 \\
Cygnus OB7 & 21       & 7 $\times$ 7      & 166.5 \\
L1642          & 44       & 2 $\times$ 2       & 9.5  \\
G93             & 61       & 2 $\times$ 2        & 6.3 \\
L183            & 41       & 2 $\times$ 2       & 9.5 \\
M31             & 84       & 3 $\times$ 3       & 6.1 \\
Polaris         & 160     & 5 $\times$ 5       & 12.3 \\
Cosmo field & 116    & 16 $\times$ 16    & 160 \\
Uranus         & 31      & 3 $\times$ 2        & 19 \\
Saturn          & 12      & 2 $\times$ 2        & 34 \\
SkyDip         & 10      &    n/a         & n/a \\
\hline
\end{tabular}
\end{table*}

The experiment successfully landed under parachute and was recovered about 350\,km east of Timmins in a dry forest area. The instrument suffered some damage, essentially to its electrical harness and optical baffle, due to collisions with trees and branches, but the science payload was successfully protected by the mechanical structure of the gondola.  Inspection of the instrument and analysis of a video recorded during the flight showed that the front baffle of the instrument deteriorated during the diurnal segment of the flight. This was caused by a defect in the thermal insulation of the baffle. It produced additional straylight which, despite substantial subtraction during data processing, limited the quality of the science data obtained during this flight. The damage to the instrument was repaired during the preparation for the second flight (see Sect.\,\ref{sec:modif_between_flights}).

\subsection{Improvements between flights}
\label{sec:modif_between_flights}

Following {\flightone}, a series of modifications were made to the instrument. In particular, the cryostat was tested in the laboratory for operations with stronger pumping on the $^4$He bath, which allowed reaching lower detector temperatures and increased the cryogenic lifetime of the cryostat from $>27$ hr to $>33.5$ hr. We also increased the size of the optical field-stop located in the cryostat, which was producing parasitic reflections affecting the polarization curves for pixels at the edges of the focal plane. We modified the thermal insulation of the optical front baffle to avoid deformation under sunlight exposure. We also decided to implement the possibility of scanning the sky at an arbitrary angle with respect to the horizon. This leaves residual atmospheric emission signal in the data that needs to be removed but can also be used to calibrate the instrument response (see Sect.\,\ref{Sec:ICS_resp}). It also allows us to obtain observations at different scan angles on the sky, which is important for optimal removal of low frequency noise through destriping, without waiting for the sky to rotate. The new scanning mode was obtained by subordinating the elevation scan speed to the cross-elevation scan speed to obtain the desired scan angle. It was tested on the ground with the actual gondola in order to check that this did not excite oscillation modes of the flight chain. We also modified the on-board computer software to allow scanning only the rectangular region around the target defined by the observer, and implemented a more flexible calibration sequence scheduler, in order to reduce overhead times. All of these changes were successfully implemented for {\flighttwo}.

\subsection{ {\flighttwo}}
\label{sec:flight_two}

The second flight was conducted from the USA-operated launch base of Alice Springs, Australia {on April 16$^{th}$ 2018}. The launch was carried out as part of a campaign led by CNES, in which three stratospheric gondolas were succesfully launched.

The flight lasted approximately 33\,hr, during which 24\,hr of scientific observations were obtained. The launch took place at 6:30 AM local time. The experiment reached ceiling altitude about 2\,hr after take-off.  The instrument reached an altitude of 39\,km, slowly decreasing to 36\,km during the first day of the flight. The altitude decreased to 31-34\,km during the night due to the lower ascensional force of the balloon, despite off-loading a fraction of the available ballast.  During the second day, the altitude rose again to reach 38-40\,km.  The focal planes temperatures evolved slightly with altitude during the ceiling period and remained in the range $\simeq304-307$\,mK and $\simeq308-312$\,mK for the {\TRANS} and {\REFLEX} focal planes respectively.  Out of the eight bolometer arrays, array \#1 ({\TRANS}), array \#3 ({\REFLEX}) and array \#5 ({\TRANS}) were not operational during {\flighttwo} for similar reasons as during
{\flightone} (see Sect.\,\ref{sec:flight_one}).

The scientific observations obtained during  {\flighttwo} are summarized in Tab.\,\ref{tab:flight2-observations}.  Apart for the Orion molecular cloud and planets, all targets observed are unique to the southern hemisphere. We mapped two regions along the inner Galactic plane near the Galactic center (L0) and near $l=30\degr$ (L30). We also obtained maps of several known molecular clouds. A large integration time was allocated to map a fraction of the Large Magellanic Cloud (LMC). We also obtained long measurements of the field observed by the {\BICEPtwo} ground experiment (\cite{BICEP2Planck_2015}) in order to attempt constraining the polarization properties of dust in a region of low Galactic foreground.

\begin{table*}[ht]
\centering
\caption{\label{tab:flight2-observations} Observations obtained during  {\flighttwo}.}
\begin{tabular}{llll}
\hline
\multicolumn{1}{c}{Source} &
\multicolumn{1}{c}{\begin{tabular}[c]{@{}c@{}}Observation Time \\
                     {[}mn{]}\end{tabular}} &
\multicolumn{1}{c}{\begin{tabular}[c]{@{}c@{}}Map size\\ {[}deg x
                     deg{]}\end{tabular}} &
\begin{tabular}[c]{@{}l@{}}Total depth\\ {[}$deg^2$/h{]}\end{tabular}  \\
\hline
L30                & 72      & 5 x 5    & 21 \\
L0                  & 32      & 2 x 5    & 18.8 \\
Orion             & 140.8 & 5 x 10  & 21.3 \\
Rho-oph        & 268.8 & 9 x 4    & 8.0 \\
Musca            & 185.6 & 2 x 3    & 1.9 \\
LMCridge       & 134.4 & 3.5 x 1 & 1.6 \\
LMCridgeBIG  & 232.5 & 4.0 x 2 & 2.0 \\
BICEP             & 290.1 & 30 x 12 & 74.5 \\
Jupiter            & 27.7   & 3 x 2    & 13.0 \\
Saturn            & 23.5   & 5 x 3.4 & 43.0 \\
Skydip           & 21.3   & n/a   & n/a\\ 
\hline
\end{tabular}
\end{table*}

The flight trajectory was eastward during most of the flight. We successfully used the two telemetry antennas located in
Alice Springs and in Longreach.  The gondola was recovered about 850\,km east of the launch site in a desert area. It was brought back to the Alice Springs base using a helicopter and a truck. The gondola and the instrument suffered no major damage from landing or recovery, which was later confirmed by a thorough inspection following the return of the instrument to France.

\section{Glitches and time constants}
\label{sec:glitches}

\noindent The {\Pilot} data are affected by `glitches', which are characterized by an abrupt deviation in the signal timeline of a bolometer followed by an exponential decay. These features can be positive or negative, and they are caused by energetic particles striking the detector absorber or the walls of the integration cavity surrounding the detectors (see \cite{Horeau+2012}). Removal of glitches from the {\Pilot} timelines is important since they do not follow the same Gaussian distribution as the detector noise, and hence produce a bias in statistical descriptions of the data, e.g. the signal mean. They are also a significant source of artifacts for steps in the {\Pilot} data processing that are performed in Fourier space. Finally, the decay following glitches can be used to constrain the time constant of individual bolometers. In this section, we give a brief description of our method for identifying glitches, a preliminary analysis of the glitch properties and the use of glitch decay to constrain detector time constants.

\subsection{Glitch identification}
\label{sec:glitches_identification}

We emphasize that our goal at this stage is to identify and suppress the most significant anomalous features in the signal timelines (i.e. those that noticeably affect the accuracy of our detector noise estimation) in an efficient, automated fashion across the entire flight, and to identify strong glitches that can be used to measure the detector time constants (Sect. \ref{sec:time_cste}). More sophisticated methods for glitch identification and signal reconstruction will be tested and applied to the {\Pilot} data in future works that present the {\Pilot} science data.

\noindent We identify glitches in the {\Pilot} timelines using the following procedure. First, we construct an estimate of the high frequency (HF, 40\,Hz$\sim1$ sample) noise by shifting the timeline for each bolometer by 1 sample, and subtracting this shifted timeline from the original timeline. We refer to the resulting timeline as the `shifted data timeline'. We then replace the value of each sample in the shifted data timeline with a local estimate of the median absolute deviation, which is calculated using adjacent samples within a rolling window of 5\,seconds. We refer to the resulting timeline as the `local HF noise timeline'. Next, we divide the shifted data timeline by the local HF noise timeline. Large (positive or negative) values in this glitch signal-to-noise ratio ($SNR$) timeline provide a preliminary list of glitch candidates. Since glitches are characterized by a sharp change followed by an exponential decay, we expect positive glitches to appear in the shifted data and glitch $SNR$ timelines as a large positive value, followed immediately by a (lower amplitude) negative value (and vice versa for negative glitches). In practice, however, this idealized characteristic signature is complicated by the presence of very strong glitches saturating the detectors for longer than one sample, detector noise (for weaker glitches) and glitches that fall exactly between two samples. {We simply disregard glitch candidates saturating the readout electronics.} We further reduce our list of preliminary glitch candidates by rejecting all candidates with $SNR<5$. We accept all glitch candidates with $SNR>10$ provided that the amplitude of the next sample has the opposite sign. For glitch candidates with $SNR \in [5,10]$, we further impose that the candidate positive (negative) value in the $SNR$ timeline should be followed by a negative (positive) value of comparable amplitude. We define `comparable amplitude' using {a softening function that requires the absolute ratio of the sample amplitudes to lie within a range that becomes less restrictive for glitch candidates with higher $SNR$,  i.e.
\begin{equation}
  1/f < \lvert \frac{g_{i}}{g_{i+1}} \lvert < f,
\end{equation}
where $f=(0.25 + 0.75 \exp(SNR/5-1)$ and $g_{i}$/$g_{i+1}$ is the ratio of the amplitudes of the consecutive samples, and we require that $g_{i}$ and $g_{i+1}$ have opposite sign.} This criterion rejects moderate significance glitch candidates ($5<SNR<10$) if they do not exhibit the characteristic positive-negative (or negative-positive).

\subsection{Glitch statistics}
\label{sec:glitch_statistics}

\begin{figure}[ht]
\begin{center}
\includegraphics[width=0.4\textwidth]{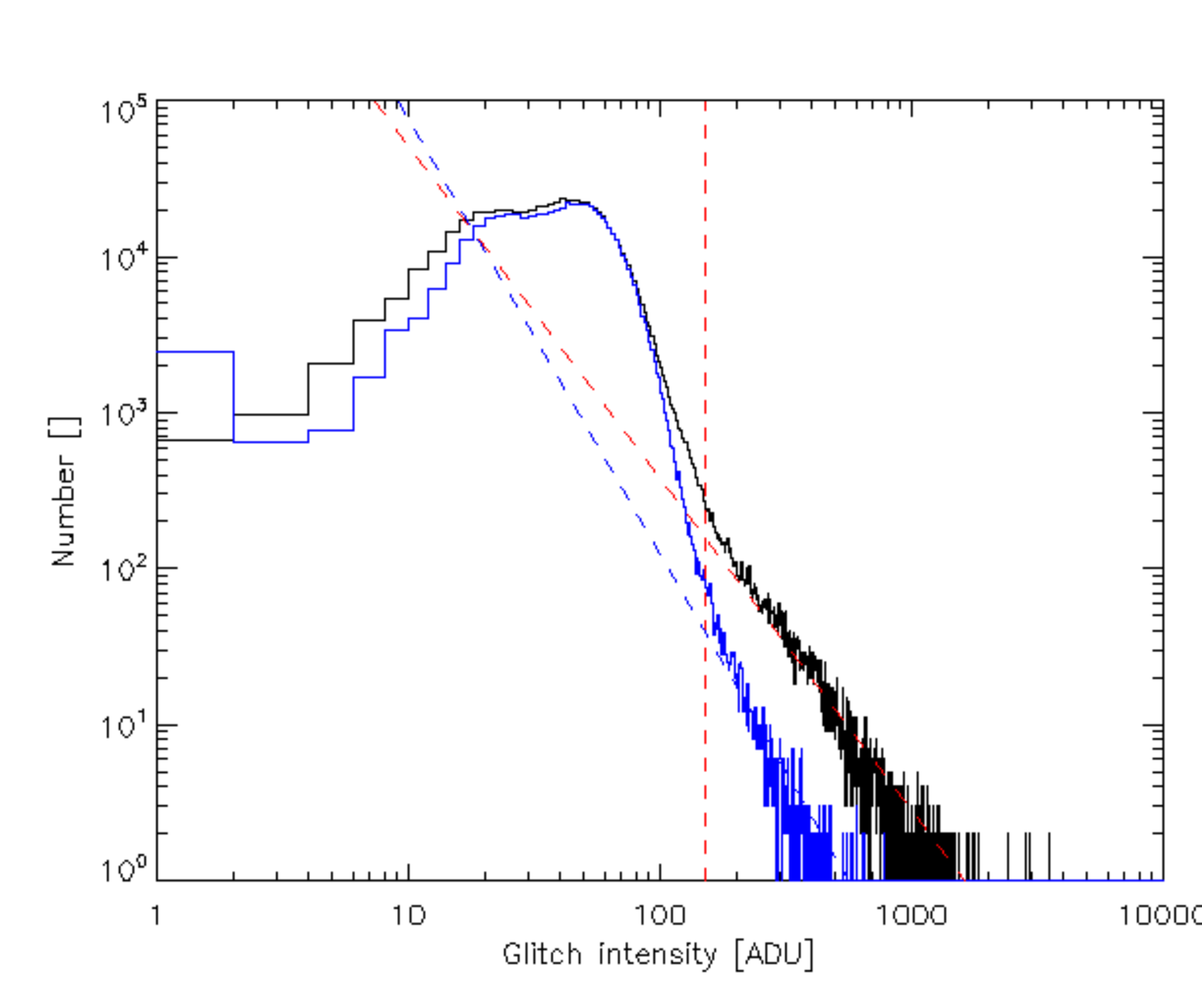}
\caption{\label{fig:glitch_histogram} Histogram of glitch intensities as measured during {\flighttwo}. The black and blue curves are for positive and negative glitches respectively. The dashed lines show fits of the distributions constrained above 150 ADU. The vertical line shows the approximate limit of 150 ADU above which true glitches dominate over noise.}
\end{center}
\end{figure}

\begin{figure}[ht]
\begin{center}
\includegraphics[width=1.0\columnwidth]{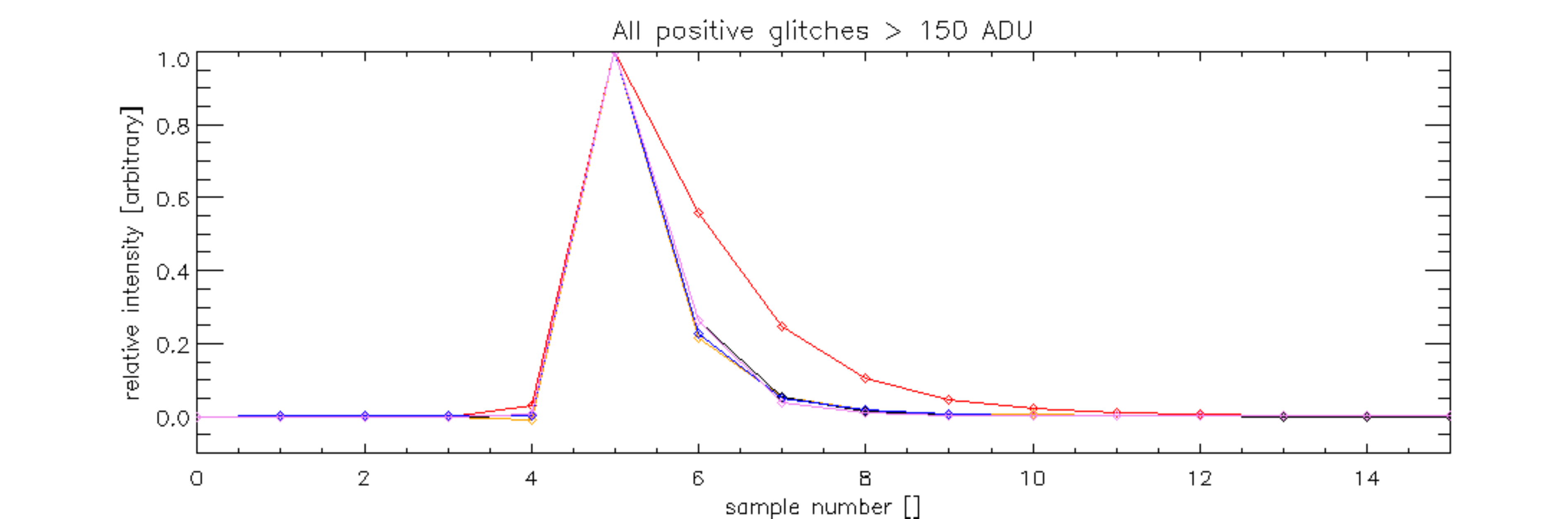}
\caption{\label{fig:glitch_stacks} Array-average of the signal following positive glitches with intensity larger than 150 ADU. The data has been co-added with the glitch position at sample number 5 and the average profiles have been normalized to their peak value. The various colors (red, orange, black, blue, violet) correspond to the various arrays (arrays \#6, 2, 4, 7 and 8 respectively).}
\end{center}
\end{figure}

The glitch intensity histogram averaged over the whole focal plane for both positive and negative glitches during {\flighttwo} is shown in Fig.\,\ref{fig:glitch_histogram}. The statistics for glitches below $\simeq 150$ ADU is affected by noise excursions also selected by the detection method.  Both positive and negative glitches are detected, with a ratio of 86.8\% for positive glitches (13.2\% for negative glitches) above an absolute glitch intensity of 150 ADU.  This ratio is similar to what has been observed for glitches on the PACS/Herschel detectors as reported by \cite{Horeau+2012}.  For glitches above 150 ADU, the glitch rate observed during {\flighttwo} is about 0.68 gli/pix/hr. This figure goes up to 2.24 gli/pix/hr for positive glitches with intensity above 100 ADU.  These values are comparable to those obtained for the PACS instrument on board Herschel by \cite{Horeau+2012}.  The glitch rate is observed to be roughly constant during {\flighttwo} and the distribution across the focal plane is mostly homogeneous.  Figure\,\ref{fig:glitch_stacks} shows the average profile of the signal decay following positive glitches for the various arrays, obtained by stacking signal around glitch locations, for glitches with intensity above 150 ADU. Above the threshold, individual pixels receive about 10 such glitches during {\flighttwo}, which allows us to constrain bolometer time constants, as described in Sect.\,\ref{sec:time_cste}. As can be seen in the figure, most arrays show similar decay profiles, except for array \#6 which shows significantly longer time constants.

\subsection{Detectors time constants}
\label{sec:time_cste}

\begin{figure*}[ht]
\begin{center}
\includegraphics[width=1.0\textwidth]{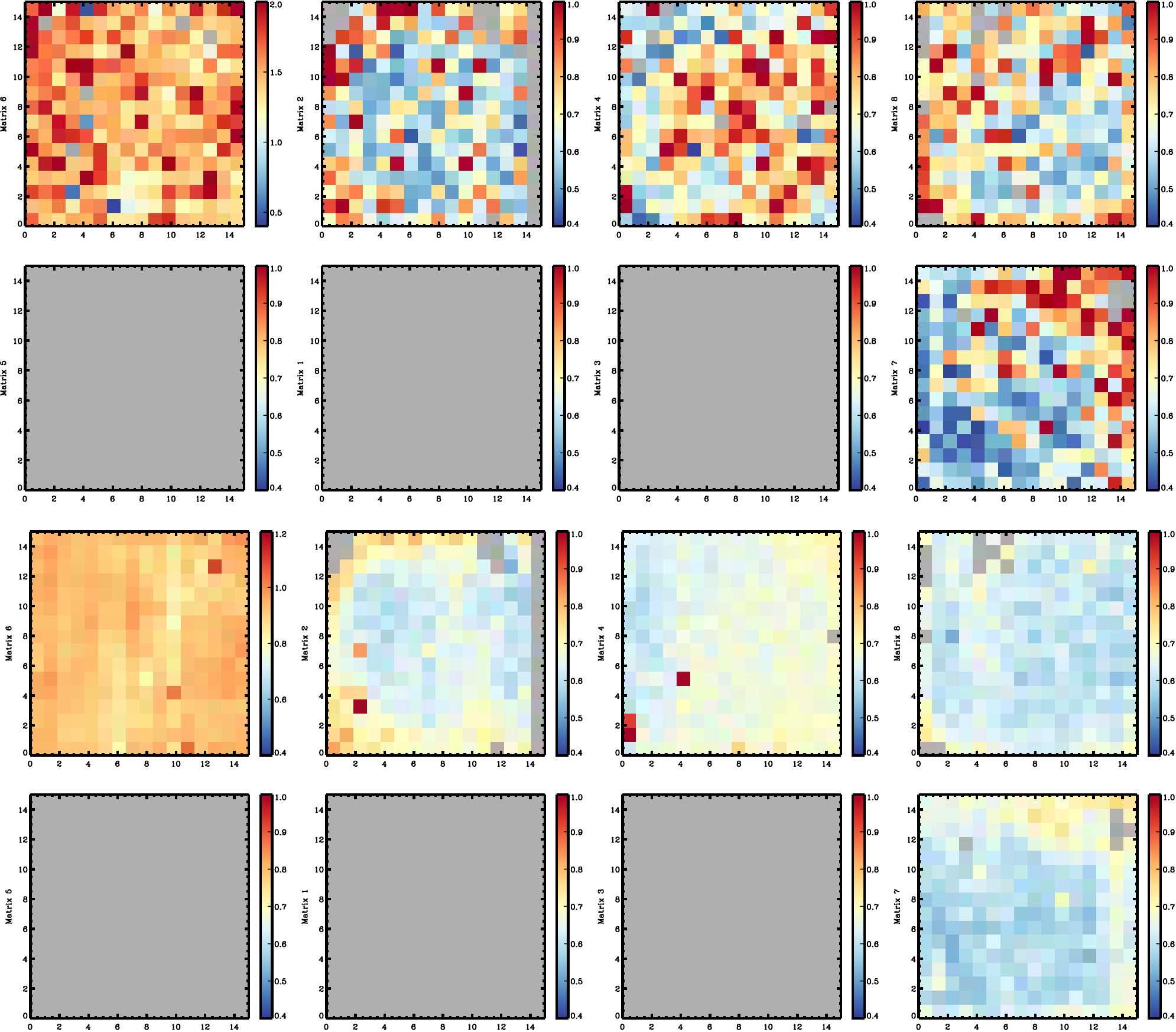}
\caption{\label{fig:tcste} Top: Focal plane distribution of the bolometer time constants as measured from glitches $\taugli$ {in units of 0.25 msec samples}. Bottom: Focal plane distribution of the bolometer time constants $\taubolo$ as measured using the {\ICS} downward transitions. The four arrays shown on the left (resp. right) belong to the {\TRANS} (resp. {\REFLEX}) focal planes, such that arrays \#6 and \#4 (or arrays \#2 and \#8) are optical conjugates. In this representation, elevation increases towards the top-left and cross-elevation increases towards the top-right corner of each focal plane. The same convention and array numbering is used for all figures of the paper.
}
\end{center}
\end{figure*}

\begin{table}[]
\centering
\begin{tabular}{c|c|c|c|c|c|c|c|c|}
\cline{2-9}
                                          & \multicolumn{3}{c|}{\TRANS} & \multicolumn{4}{c|}{\REFLEX} & ALL  \\ \hline
\multicolumn{1}{|c|}{Array}               & 2       & 6       & Avg    & 7     & 4     & 8    & Avg  & Avg  \\ \hline
\multicolumn{1}{|c|}{$\taugli$}   & 0.61    & 1.51    & 1.06   & 0.69  & 0.74  & 0.68 & 0.70 & 0.85 \\ \hline
\multicolumn{1}{|c|}{$\taubolo$} & 0.61    & 0.89    & 0.75   & 0.58  & 0.63  & 0.58 & 0.60 & 0.66 \\ \hline
\end{tabular}
\caption{Array-averaged values of the detector time-constants as measured using glitches and using {\ICS} downward decay assuming an intrinsic {\ICS} time constant of $\tauics$=11\,ms (see text).}
\label{average_time_constant}
\end{table}

During the calibration sequences with the {\ICS}, the bolometric signal following each {switching off} of the source gradually decreases with a characteristic time $\taucal$.  This progressive decay results from the convolution of the square pulse controlled by the current injected into the {\ICS} with the transfer function of the {\ICS} and that of the bolometers. A similar decrease follows the absorption of high energy particles (glitches) by the detectors, which can also be used to determine the response of individual detectors.

In the following, we assume that the transfer functions of the {\ICS} and the bolometers can be described by an exponential decay of the form $e^{-t/\tau}$, where $\tau$ is the time constant.  We denote the time constants of the {\ICS} as $\tauics$, and the time constants of the individual bolometers as $\taubolo$.

Because the glitch rate is low, determining $\taubolo$ from glitch transients alone would lead to a noisy determination. Instead, we use a combination of the glitch and {\ICS} transients described below. The array-averaged values determined for each array are summarized in Tab.\,\ref{average_time_constant}.

In a first step, we estimate a detector's time constant using glitches, which we call $\taugli$.  We averaged the measurements over 15 samples following positive glitches with intensity larger than 200\,ADU.  The average values derived for each array are given in Tab.\,\ref{average_time_constant} and the focal plane distribution of $\taugli$ is shown on Fig.\,\ref{fig:tcste}. The time constant derived using glitches is around 0.7\,samples (17.5\,ms) for most arrays, except for array \#6, for which it is around 1.5\,samples (37.5\,ms). {Note that this procedure does not correct for the broadening of the average profile due to the random arrival time of glitches.}

In a second step, we measure the time constant of the {\ICS} using calibration sequences.  We stack all decays of the signal after {turning off} the {\ICS} and compute the corresponding average {\ICS} downward profile viewed by each pixel.  We then deconvolve these average profiles in Fourier space, using the transfer function $e^{(-t/ \taugli )}$. We then average these profiles over pixels and fit the average profile with an exponential function and derive a value of $\tauics$ of 0.41\,samples (10.25\,ms).  This can be compared to the value of 9.23\,ms measured by \cite{Hargrave2006} for a device similar to our source and operated under similar conditions. {Note that in this procedure, we take advantage of the fact that individual lines of detectors of a given array are actually read-out with a known time-delay of $1/16^{th}$ of the readout frequency to increase our knowledge of the {\ICS} switching off time, therefore mitigating the broadening of the average profile.}

In the last step, we determine the final values of $\taubolo$.  We proceed in the same way as for the measurement of $\tauics$, except that we use here the above value of $\tauics$ as a parameter for the deconvolution kernel and fit the downward profile of each pixel in order to derive the values of $\taubolo$ for each pixel.  The array-averaged values of $\taubolo$ are given in Tab.\,\ref{average_time_constant} and are of the order of 0.60\,samples (15\,ms).  On average, array \#6 is slower than other arrays, with an average $\taubolo$ of 0.89\,samples (22.25\,ms).  The focal plane distribution of $\taubolo$ is shown on Fig.\,\ref{fig:tcste}.

\section{Point spread function}
\label{sec:psf}

All optical elements in the {\Pilot} instrument except the  primary mirror M1, are cooled below 3K inside the cryostat.  The overall optical quality of the system is therefore sensitive to the positioning of the primary mirror with respect to the cryostat. External conditions can modify the relative position of the optics during flight, in particular thermo-elastic and bending under gravity effects in the mechanical structure holding the primary mirror. To minimize thermal effects, the pre-flight optical alignment was based  on the flight thermal modeling, and thermo-elastic analysis of the instrument and inflight temperature predictions \cite{longval2016pilot}. Checking the optical quality using inflight measurements is therefore of particular importance.

During each flight of the instrument, we observed planets, which can be considered as point sources at the resolution of {\Pilot}. These observations can therefore be used to assess the optical quality through a measurement of the instrument Point Spread Function (PSF).  Planet maps were constructed with a pixel size of 0.1$^\prime$ for each detector array and for each scan through the planet (referred to as planet crossings). The data have been corrected for the responses calculated on the skydips, as described in Sect.\,\ref{Sec:ICS_resp} and corrected for the effect of the detector time constant through deconvolution using the values derived in Sect.\,\ref{sec:time_cste}.

\begin{figure}[ht]
\begin{center}
\includegraphics[width=1.0\columnwidth]{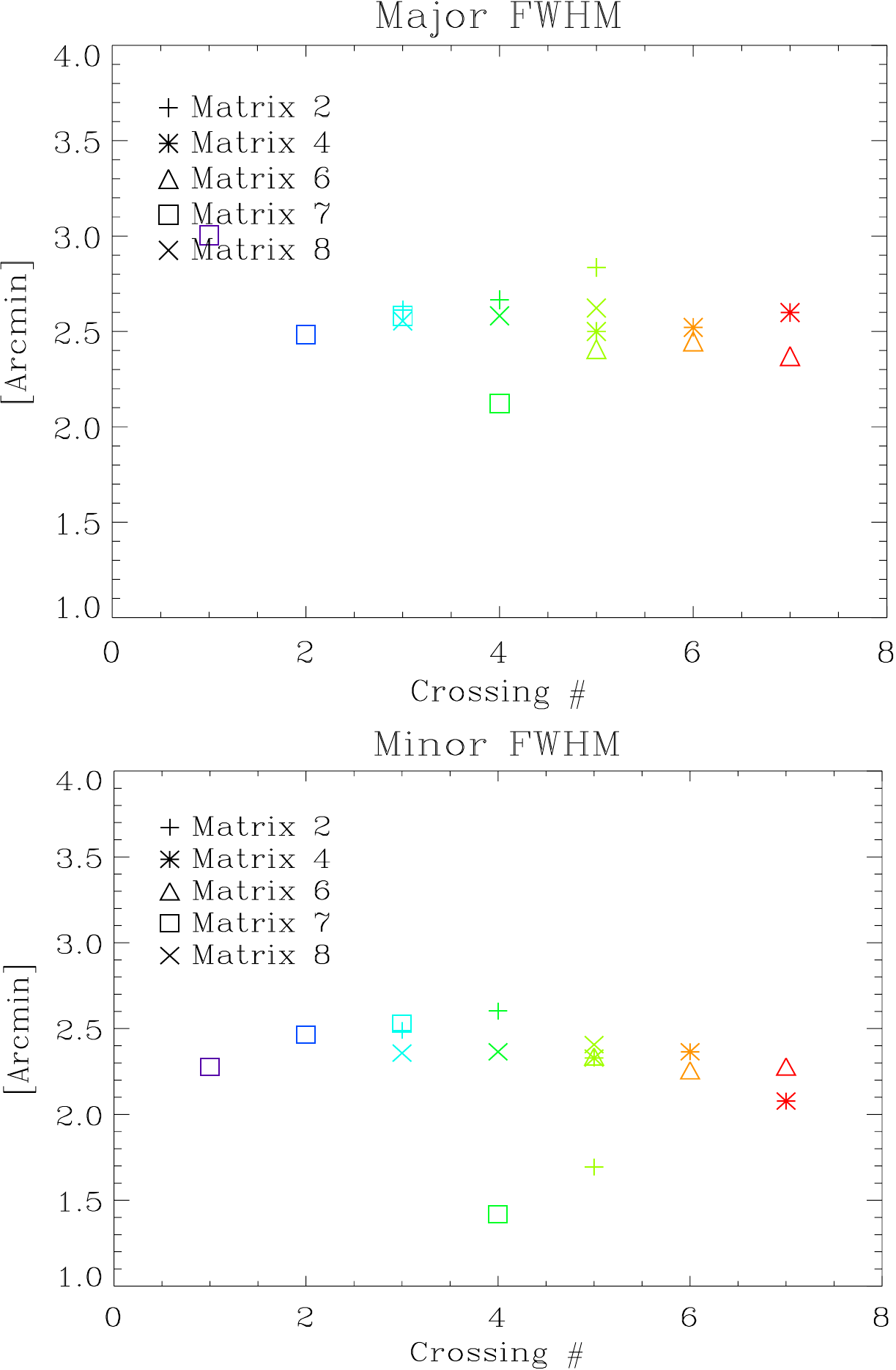}
\caption{The top and bottom panels show the evolution of the PSF FWHM measured on Jupiter along the minor and major axes respectively during {\flighttwo}, as a function of the planet crossing number. The different colors correspond to the different crossings of the planet on the focal plane and the different symbols correspond to individual detector arrays.}
\label{fig:psf-bestfit_flight2}
\end{center}
\end{figure}

The parameters of the PSF were deduced using an elliptical Gaussian fit applied to the individual crossing images.
Figure\,\ref{fig:psf-bestfit_flight2} shows the values obtained for the major and minor axes dimensions of the PSF obtained on Jupiter. {Note that the outliers in the plot do not reflect actual changes of the PSF FWHM, but correspond to situations where the PSF image for a given crossing and array falls close to the image edge, affecting the 2D Gaussian fit.} We obtain an average full-width half maximum (FWHM) size of 2.25$^\prime$ $\pm$ 0.15 $^\prime$, taking into account the apparent size of the planet (44.2$^{\prime\prime}$). The uncertainties were derived from the statistics over the various planet crossings.

\begin{figure}[ht]
\begin{center}
\includegraphics[width=1.0\columnwidth]{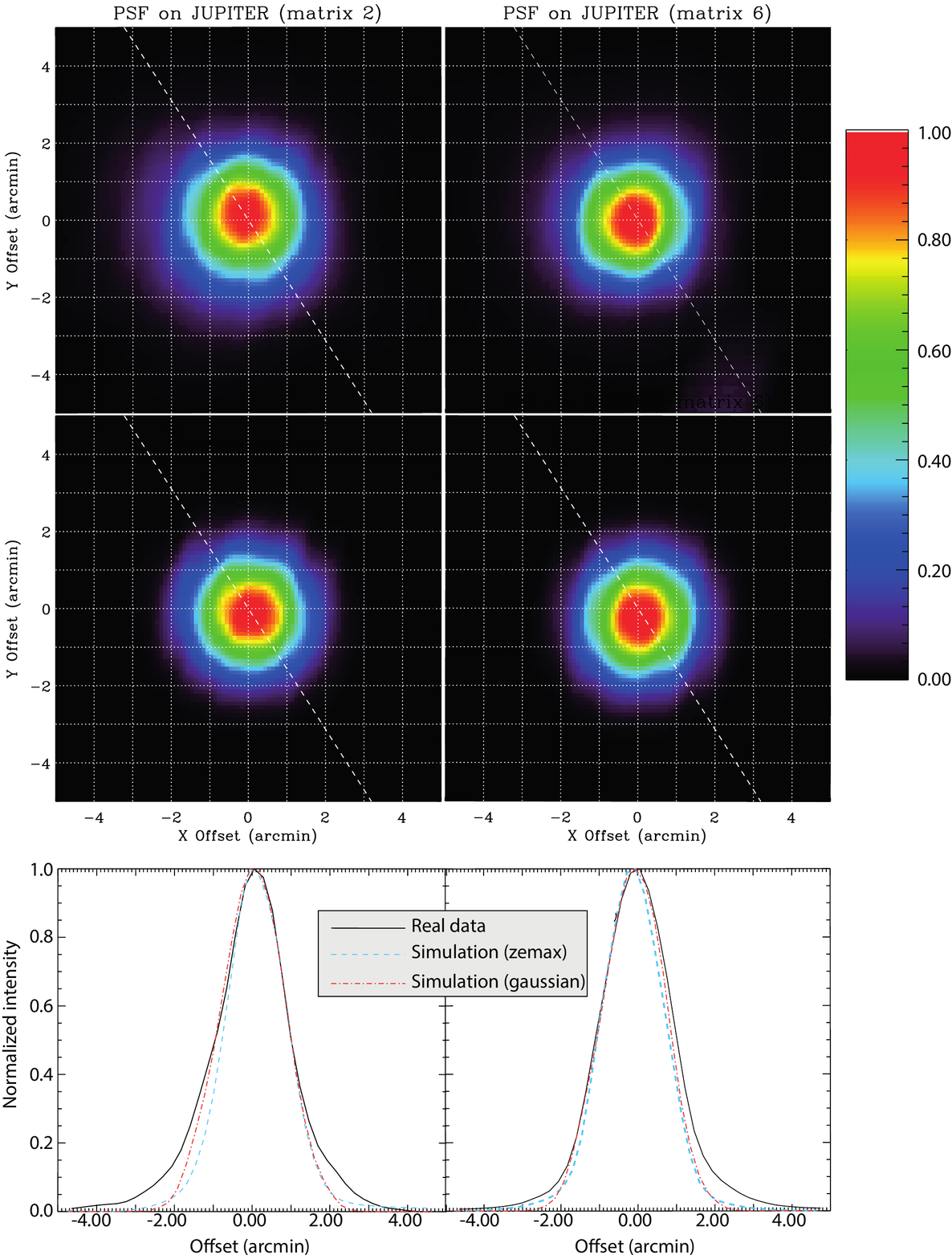}
\caption{Top: Images of Jupiter obtained with array \#2 (left) and
6 (right) during one of the observing sequence of the planet. Middle: Map obtained on array \#2 (left) and \#6 (right) from signal simulated using the actual pointing (see text). The scan direction is shown by the dashed white line. Bottom: Circular average profile of the PSF measured on array \#2 (left) and array \#6 (right). The solid black curve corresponds to measurements from the observations on Jupiter during {\flighttwo}. The dashed blue curve is obtained from a simulation (see text) using an input PSF from the modelling of the optical system with the zemax software \cite{Engel_thesis_2012}. The red dash-dotted curve is obtained from the same simulation using a PSF approximated with a Gaussian function. Each PSF has been normalized by its maximum value.}
\label{fig:psf-flight2}
\end{center}
\end{figure}

Figure\,\ref{fig:psf-flight2} shows the normalized average total intensity image over all crossings of Jupiter obtained with array \#2 and array \#6 during one observation of {\flighttwo}.  The PSF shows no particular elongation along the scan direction, indicating accurate accounting for the detector time constants.  However, we can see that the PSF images have a `boxy' shape, which likely reflects that they result from the convolution of the optical PSF with a square pixel.

To understand the origin of the ``box effect'' which may be induced by the convolution of the PSF and the size of the bolometers pixels, we simulated timelines calculated as the integral of the flux received by a square detector with a pixel size matching that of {\PILOT} (1.4$^\prime$) observing a Gaussian diffraction-limited optical PSF with the expected size for {\PILOT} ($\approx 1.4^\prime$) centered at the predicted position of the planet. The resulting map is compared to the observed PSF in Fig.\,\ref{fig:psf-flight2}. The simulated PSF also shows the observed ``box effect''. This confirms the hypothesis that they result from the convolution of the optical PSF with a square pixel. Applying the same analysis to this simulation as to the data, we measure a FWHM of 2.31 $^\prime$ $\pm$ 0.07 $^\prime$.  The value obtained in the simulation is therefore consistent with the value measured in-flight.

Figure\,\ref{fig:psf-flight2} also shows the circular average profile of the PSF measured on array \#2 and array \#6 and the comparison with the simulations.  The profiles are quite similar, except for the presence of more intense PSF wings in the measurements, an effect that was also observed during ground tests.

\section{Instrumental Background }
\label{Sec:Inst_bkd}

We estimated the background level inflight using a combination of calibration measurements obtained at ceiling altitude and during ground tests. These calibration measurements involve, firstly, a short sequence of data recorded with different detector settings, from which we obtain an empirical conversion between the signal measured in ADU and its equivalent in Volts. This procedure is relatively quick, and was repeated during ground tests and at the beginning of both \PILOT\ flights. A second -- much longer -- ground calibration measurement procedure is required to establish the relation between the output voltage of each bolometer and the intensity of the background incident on the detectors. These measurements were performed during tests in front of a controlled black body at cryogenic temperature in 2012 and 2016.

\subsection{Background Level}

\begin{figure*}[ht]
\begin{center}
\includegraphics[width=1.0\textwidth]{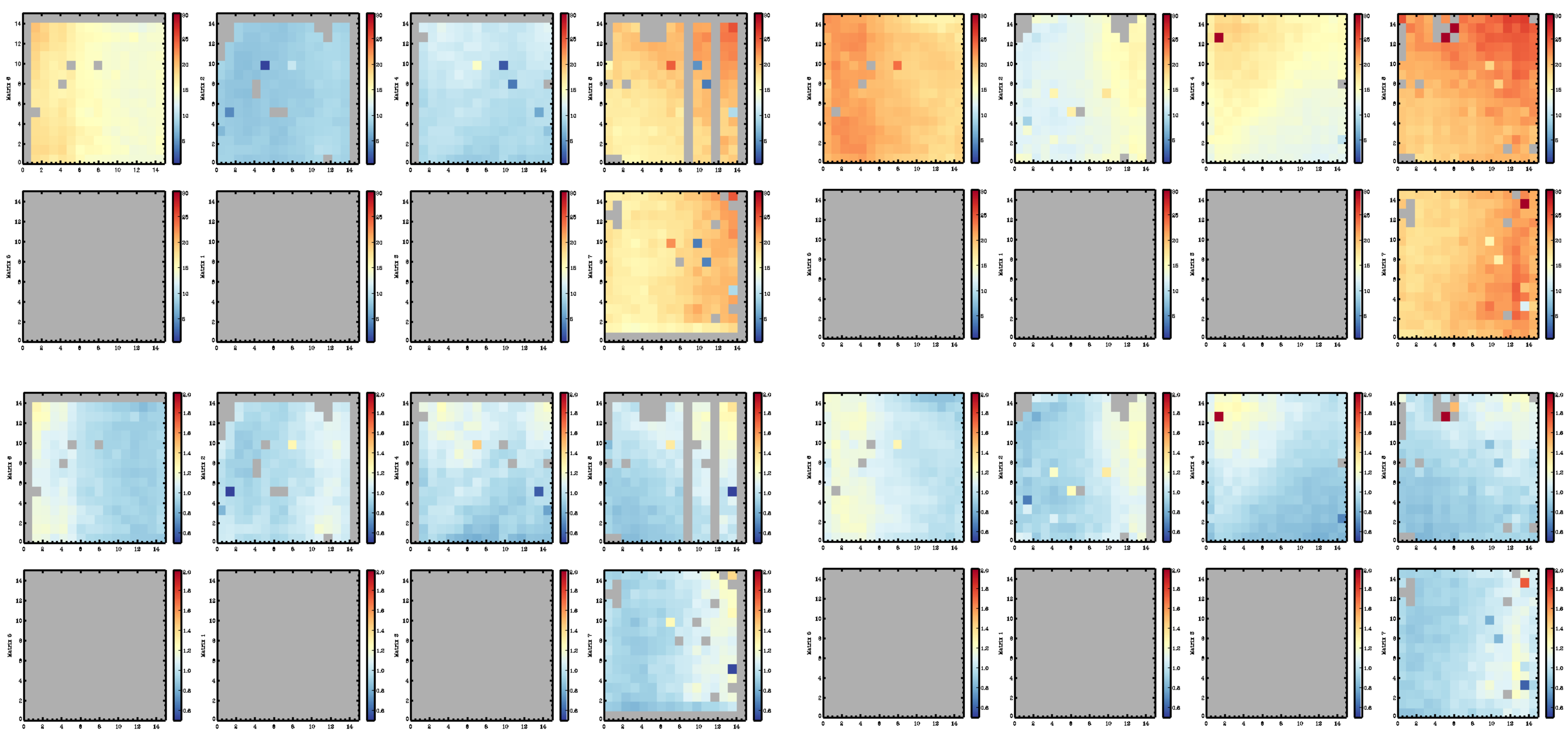}
\caption{\label{fig:background-image} Example focal plane images of the background derived during {\flightone} (left) and {\flighttwo} (right), each from $\sim30$\,seconds of observation. The top row shows an estimate for the absolute background level, the bottom row shows the background level normalized by the average on each array. The representation of the focal plane is the same as in Fig.\,\ref{fig:tcste}.
}
\end{center}
\end{figure*}

Figure\,\ref{fig:background-image} shows examples of the background image obtained during {\flightone} and {\flighttwo} using the technique described above.

The typical background value measured during {\flightone} and {\flighttwo} were in the range 13-16\,pW per pixel towards the center of the focal plane.  These absolute values are quite uncertain due to the limited sampling of the calibration measurements, which could only be conducted for a finite set of detector settings and incident background levels and the difficulties in interpolating for the inflight settings. As seen in the top row of Fig.\,\ref{fig:background-image}, this is evident from the unphysical offsets between the array-averaged background levels that we infer (i.e. the average background on arrays \#2 and \#4 is lower than on the other arrays). Nevertheless, it seems likely that the background during both {\PILOT} flights was higher than the values predicted using our instrument photometric model of 5.7\,pW per pixel at the center of the focal plane (see \cite{Bernard_etal2016}). Variations in the background level were monitored throughout each flight, and were found to be relatively stable except for variations related to residual atmospheric emission and {\HWP} position, due to the background polarization (see Sect.\,\ref{sub_sec:pol_Inst_bkd}). The background distribution in the focal plane is shown in the bottom row of Fig.\,\ref{fig:background-image}, where we have normalised the pixel values by the average background level measured on each array. The shape of the background follows a similar distribution as observed during ground calibrations, with values rising by about a factor of two from the center to the corners of the focal plane. This distribution is explained by the absorption in lens L2, located just in front of the focal planes.

\subsection{Background polarization}
\label{sub_sec:pol_Inst_bkd}

\begin{figure}[ht]
\begin{center}
\includegraphics[width=1.0\columnwidth]{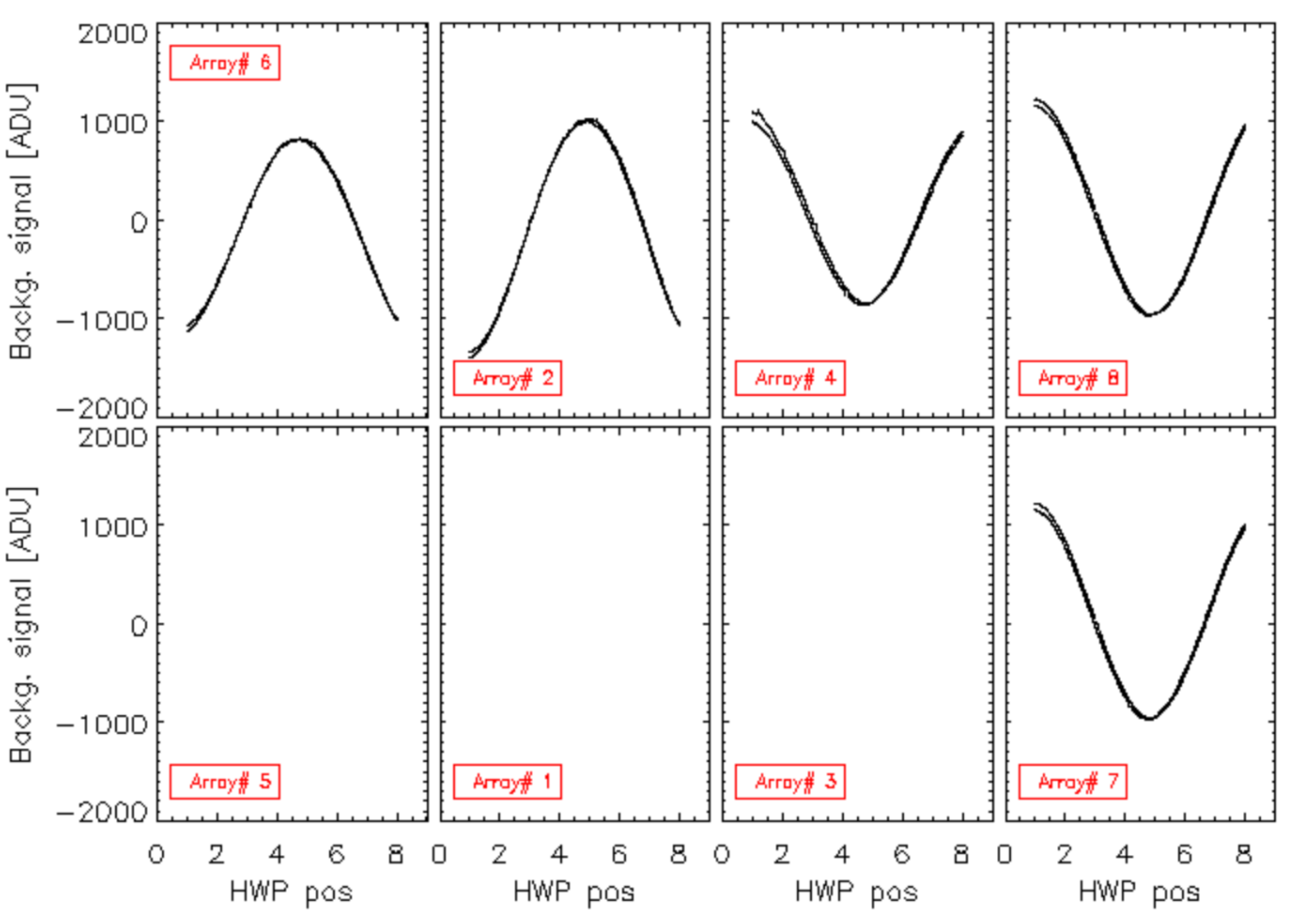}
\caption{\label{fig:backgroundpolarization_flight2} Variation of the background signal as a function of the {\HWP} for all arrays during one observation of {\flighttwo}. The sine curve with opposite phase on the {\TRANS} and {\REFLEX} arrays is due to the polarization of the instrumental background emission.}
\end{center}
\end{figure}

\begin{figure}[ht]
\begin{center}
\includegraphics[width=1.0\columnwidth]{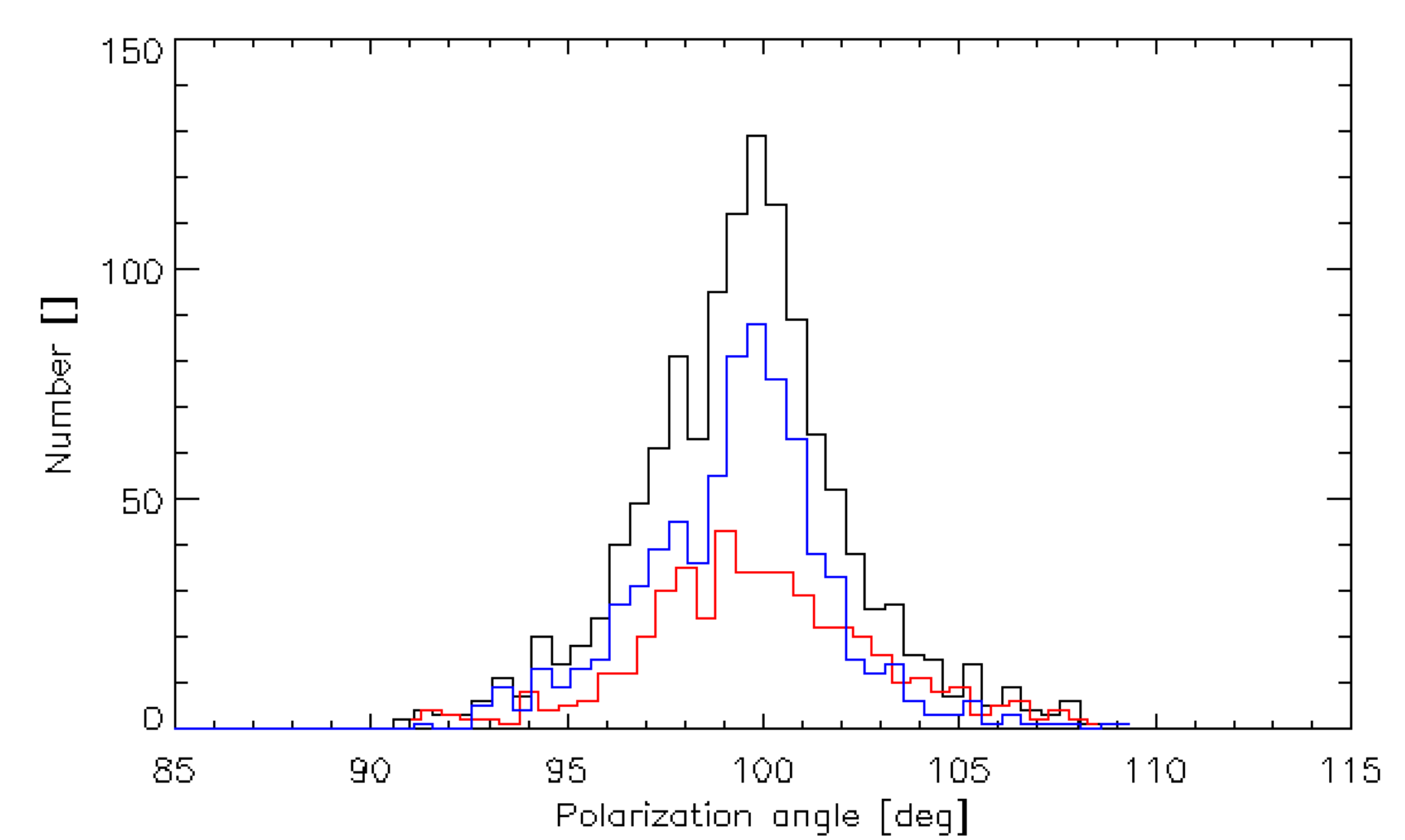}
\caption{\label{fig:psi_histogram} Histogram of {the measured polarization angle $\Psi$ for the instrumental background}. The black, red and blue lines show the curves for all pixels and the {\TRANS} and {\REFLEX} pixels respectively.}
\end{center}
\end{figure}

Figure\,\ref{fig:backgroundpolarization_flight2} shows the variations of the signal observed on each array as a function of {\HWP} position, measured during {\flighttwo}. During these observations, the {\HWP} was moved successively from positions 1 to 8 and then back to position 1, {therefore exploring angles between about $-45\degr$ and $+45\degr$ from the vertical direction.}
The signal adopts a sinusoidal shape {in phase opposition between the {\TRANS} and the {\REFLEX} arrays}. We used this data to derive the polarization properties of the background. The signal amplitude varies by about $2.10^{3}$ ADU, which corresponds to 0.5\,pW, corresponding to 2.5\% of the measured background level. The maximum signal on the {\TRANS} array (or minimum signal on the {\REFLEX} arrays), is at around {\HWP} position 5, which corresponds to the fast axis of the {\HWP} being roughly vertical in the instrument restframe. Given the orientation of the polarizer in the instrument, this implies a polarization direction roughly horizontal. Figure\,\ref{fig:psi_histogram} shows the histogram of the polarization angle in the instrument restframe, deduced using a fit of the data using Eq.\,\ref{eq:pol_measure_easy_Stokes}, with $\omega$ being the {\HWP} angle in the instrument reference frame, increasing counterclockwise when looking at the sky. The polarization angle values are well peaked around 100$\degr$, indicating that the polarization direction is roughly horizontal and constant over the focal plane. Very similar polarization angles and fractions were observed during {\flightone}.

Note that a similar polarization of the background was observed during ground calibrations, as reported in \cite{Misawa_etal2017}. Within the large uncertainties on the absolute value of the background level, the polarization fraction on the ground was similar to that observed in flight. However, the polarization angle is markedly different, with horizontal polarization in-flight and at around $\psi=-45 \degr$ during ground calibration. We currently attribute this rotation to a different origin of the background in the two situations. While the inflight background is mostly due to the instrument, the background measured during ground calibrations is dominated by the room atmosphere in the few first centimeters in front of the entrance window of the cryostat. The fact that the angles are strongly rotated indicates that polarization probably arises from propagation of the unpolarized background through the instrument, with a differential rotation depending on where in the instrument the background originates.

\section{Detector response}
\label{Sec:resp}

The response of bolometers, which measures their ability to convert flux variation into an electrical signal variations, varies with their temperature and with the optical background that they receive.  It is important to accurately quantify these variations in order to calibrate the data. This is especially true for polarization measurements with the {\PILOT} instrument, which rely on combining data taken by different detectors at different times and with different {\HWP} positions. We use our Internal Calibration Source ({\ICS}) to precisely measure the time variations of the response, and we use the variations of the residual atmospheric signal in order to measure the spatial variations of the responses (or response flat-field) in the focal plane.

\subsection{Response time variations}
\label{Sec:ICS_resp}

The {\ICS} source is turned on at regular intervals during flight, at the end of individual mapping scans and occasionally during instrument manoeuvres. Calibration sequences typically account for $\simeq 5$ ON-OFF cycles. The source is driven with a square modulated current with a period of $\simeq$1\,sec, and the current ($I$) and voltage ($V$) are recorded continuously.  The ground calibration tests have shown that the {\ICS} optical flux measured by the detectors is proportional to the electrical power $P=V \times I$ dissipated in the source
(see \cite{Misawa_etal2017}).

In practice, in order to correct for amplitude drifts unrelated to the {\ICS} signal, we subtract drifts due to the residual atmospheric emission in each timeline, using the correlation with pointing elevation and apply a {high-pass} filter to the bolometer signal timelines. In order to mitigate the effects associated with the time constants of the bolometers, a few data samples at the beginning and end of each {\ICS} sequence are discarded. We only consider {\ICS} sequences when the {\HWP} is not moving and discard truncated {\ICS} sequences with less than 4 ON-OFF cycles. We then compute the response of each bolometer to the {\ICS} signal for a given calibration sequence as
\begin{equation}
\RespICS =\frac{\DeltaICS}{\langle \RICS (\langle \IICSON \rangle^2 -
  \langle \IICSOFF \rangle^2)\rangle}  \times \RICSref \times \IICSref^2,
\label{eq:ics_resp}
\end{equation}
where $\DeltaICS$ is the observed average signal difference in ADU between the ON and OFF states of the source, $\RICS$ is the source impedance and $\langle \IICSON \rangle$ and $\langle\IICSOFF \rangle$ are the time-averaged currents during the ON and OFF periods of the calibration sequence respectively. $\RICSref=300$ $\rm \Omega$ and $\IICSref=1.8$ mA are the reference impedance and current values used for normalization. $\RICS$ is measured as the time-average of $\VICS /\IICS$ over the ON periods.

\begin{figure*}[ht]
\begin{center}
\includegraphics[width=1.0\textwidth]{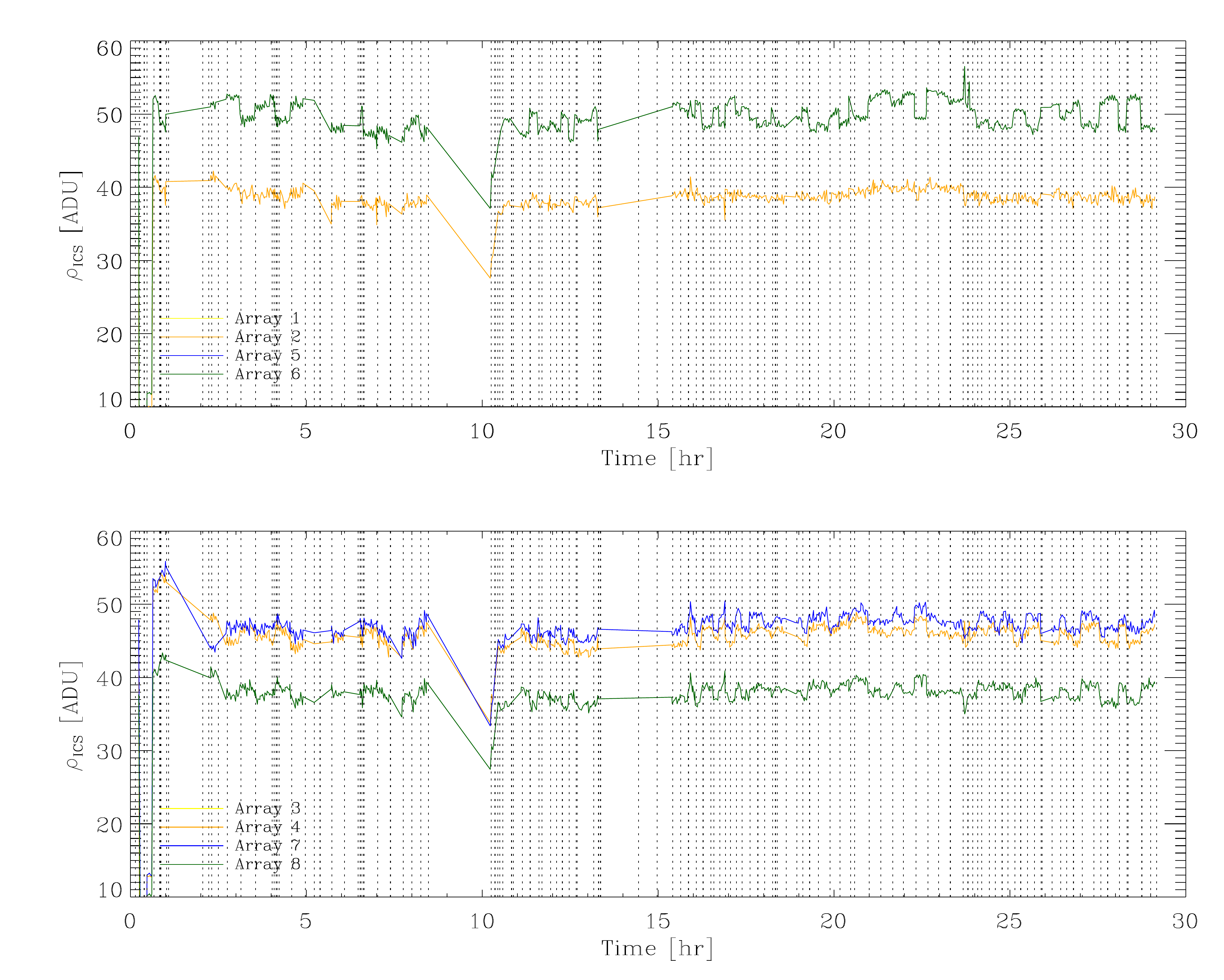}
\caption{\label{fig:response-flight2} Top: Time variations of the array-averaged detector response to the {\ICS} for arrays on the {\TRANS} focal plane (arrays \#6 and \#2) during {\flighttwo}. Bottom: Same for the {\REFLEX} focal plane (arrays \#4, \#7 and \#8).  The vertical dashed lines show the boundaries between different observations.}
\end{center}
\end{figure*}

Figure\,\ref{fig:response-flight2} shows the time variations of the array-averaged detector response to the {\ICS} signal for individual arrays during {\flighttwo}. In general, array \#6 has the best response and is $\simeq 25$\% more responsive than arrays \#2 and \#8. The variations in the array-averaged response with time are about 10\% for all arrays.  Step-like variations are clearly seen in the response of both the {\TRANS} and the {\REFLEX} focal plane arrays. These variations are mostly caused by variations of the background level on the various detectors between individual observations. Some of these variations are due to observation elevation, which changes the intensity of the residual atmospheric emission and therefore the optical background in the same way on both focal planes.  Some variations are caused by observing with different {\HWP} angles, which, due to the polarized instrumental background (see Sec.\, \ref{sub_sec:pol_Inst_bkd}), changes the optical background in opposition on the two focal planes, causing reversed variations of the detector responses.

\begin{figure*}[ht]
\begin{center}
\includegraphics[width=1.0\textwidth]{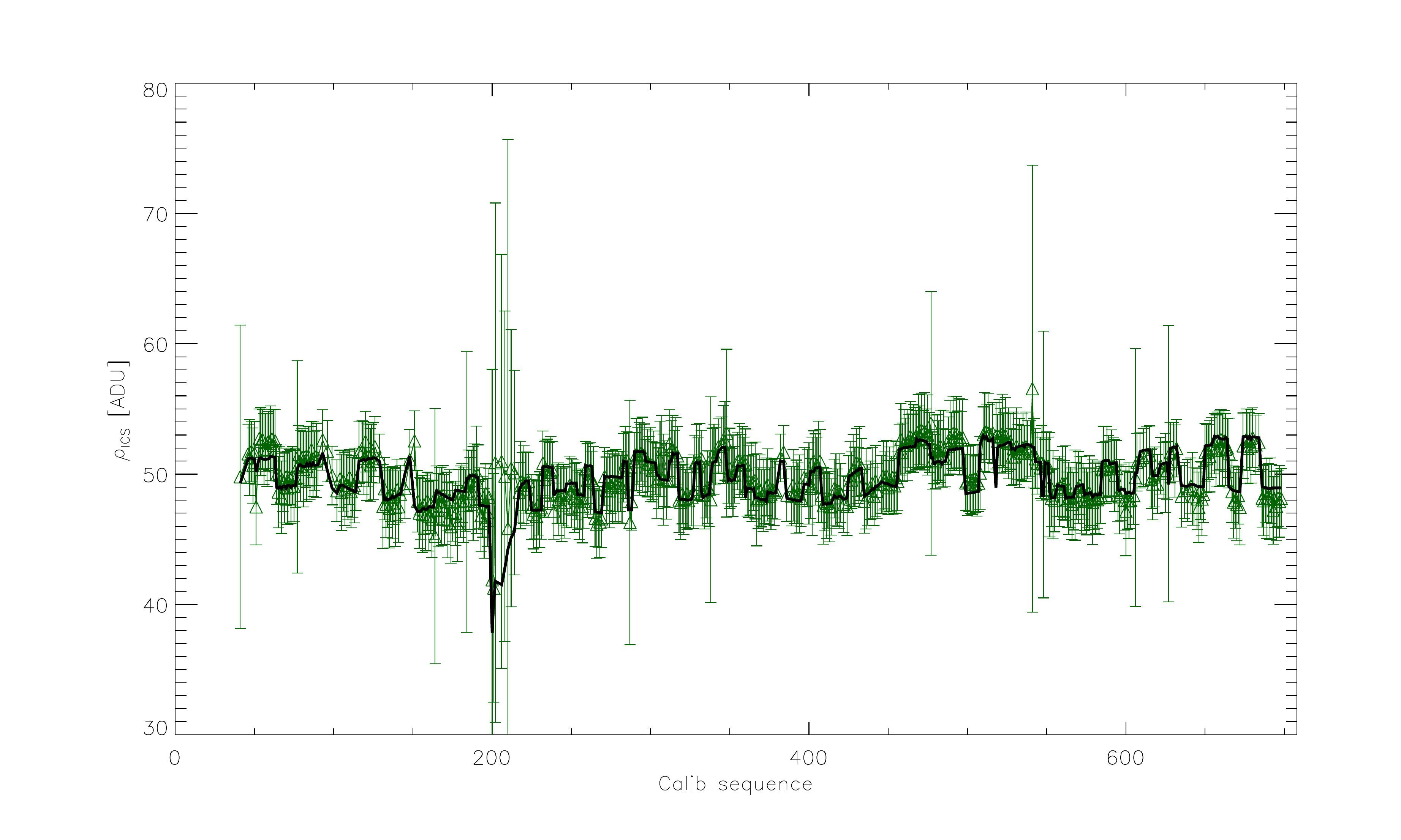}
\caption{\label{fig:ics_resp_mod_arr6_flight2} Array-averaged {\ICS} response for array \#6 as a function of the calibration sequence number (green) compared to the predictions of the parametric model described in Sec.\,\ref{Sec:ICS_resp} (black).}
\end{center}
\end{figure*}

\begin{table}
\begin{center}
{\footnotesize 
\begin{tabular}{lll}
\hline\hline
Template & c.c.  & origin \\
\hline
$cos(4 \times \omega)$  &   -0.62 & inst. backg. polarization \\
$sin(4 \times \omega)$  &   -0.54 & inst. backg. polarization \\
$90\degr$-elevation     &   -0.39 & atmospheric backg. \\
$\Tfen$                 &   -0.33 & inst. backg. \\
$\TFP$                  &   -0.32  & FP temperature \\
$\TMirror$              &   -0.24  & inst. backg.\\
altitude                &    0.22  & atmospheric backg. \\
\hline
\end{tabular}
}
\end{center}
\caption{
\label{tab:Ics_resp_mod_coeff}
Description of the various components included in the array-averaged bolometer response model for array \#6 shown in Fig.\,\ref{fig:ics_resp_mod_arr6_flight2} and defined in Sec.\,\ref{Sec:ICS_resp}. The first column gives the house-keeping template name, the second column gives the {Correlation coefficients} between the response and each component, the last column is a description of the physical origin.}
\end{table}

{In order to understand the origin of the response time variations, we performed a linear regression of $\RespICS$ with several house-keeping (HK) data recorded in the telemetry.
The array-averaged bolometer response is modeled as:
\begin{equation}
\RespICS ^{model}=\sum_{i=1}^{9} \alpha_i \times \frac{xp_i }{\langle xp_i \rangle} + c,
\end{equation}
where $xp_i$ is the template of a given house-keeping parameter averaged over the calibration sequence, $\langle \rangle$ designates averaging over all calibration sequences and $c$ is a constant.
We included in the fit HK parameters likely to capture the effect of the instrumental background polarization such as the HWP angle position, the background intensity through several temperatures taken near optical elements, the atmospheric residual emission though pointing elevation and gondola altitude, or measuring the focal planes temperatures. We also considered other HK parameters not related to expected cause of response variations, allowing for instance for a temporal slow drift of the response.}

{In order to assess which HK parameters have a significant impact on the fit, we computed the F-test and the p-value statistics and rejected HK parameters with p-value lower than $\alpha=5\%$. We found that the only relevant HK parameters are those listed  in Tab.\,\ref{tab:Ics_resp_mod_coeff} in order of decreasing correlation coefficient. The main correlation is with the {\HWP} angle, indicating a strong dependence to the background level through its polarization.
The atmospheric parameters (elevation and altitude) are also important, showing a significant impact of the residual atmospheric emission on the detector response. The correlation of the response variations with the temperatures of the instrument (such as the primary mirror temperature $\TMirror$ and the entrance window temperature $\Tfen$) also shows the impact of the instrumental background and explains the low frequency variations observed in Fig.\,\ref{fig:ics_resp_mod_arr6_flight2}. These results show that, within the measurement uncertainties, the detector response variations are mostly caused by variations of the instrumental background level, and to a lesser extent, to variations of the focal plane temperature, with no other significant contribution.}

Figure\,\ref{fig:ics_resp_mod_arr6_flight2} compares the prediction of the above model with the time evolution of the average bolometer response to the {\ICS} signal for array \#6. The uncertainties shown on the figure were computed from the statistics of the individual ON-OFF measurements in each calibration sequence.  In general, the model describes the data with good accuracy over the whole flight for all the arrays{, with a reduced $\chi^2$ of 0.1}. The median difference between the model and the data is around 2\%. A few exceptions for which the model does not match the data are visible in Fig.\, \ref{fig:ics_resp_mod_arr6_flight2}, which are associated to a larger uncertainty in the response estimate. These observations, which are evident around calibration sequence 200 in the figure, occurred just after the recycling of the $^3$He fridge, which caused large temperature fluctuations of the 300 mK stage.

\subsection{Response spatial variations}
\label{Sec:atmo_resp}

\begin{figure*}[ht]
\begin{center}
\includegraphics[width=1\textwidth]{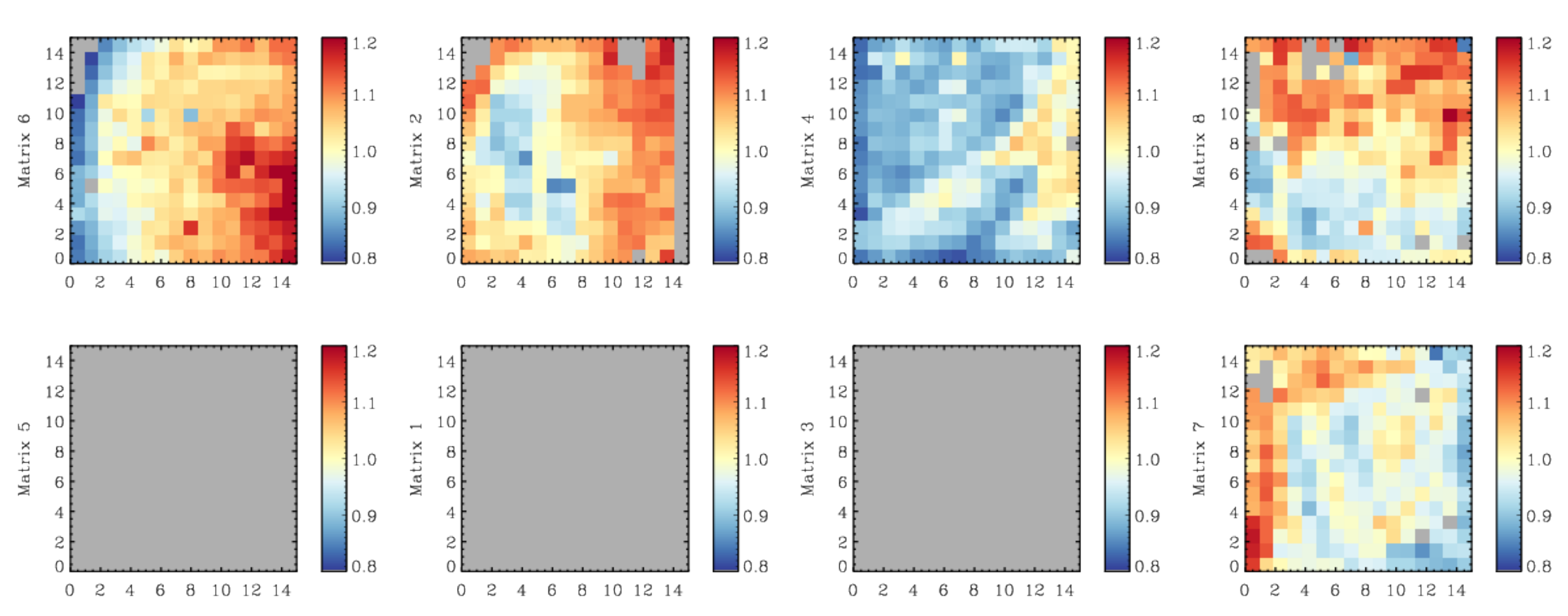}
\caption{Focal plane map of the mean response computed on the residual atmospheric signal during skydip observations of  {\flighttwo}. The response map has been divided by its average over valid detectors so that it has a mean equal to unity. The relative dispersion between these values and the one computed during the rest of the flight (see Fig.~\ref{fig:response_map_scans}) is 0.9 \%.}
\label{fig:response_map_skydips}
\end{center}
\end{figure*}

\begin{figure*}[ht]
\begin{center}
\includegraphics[width=1\textwidth]{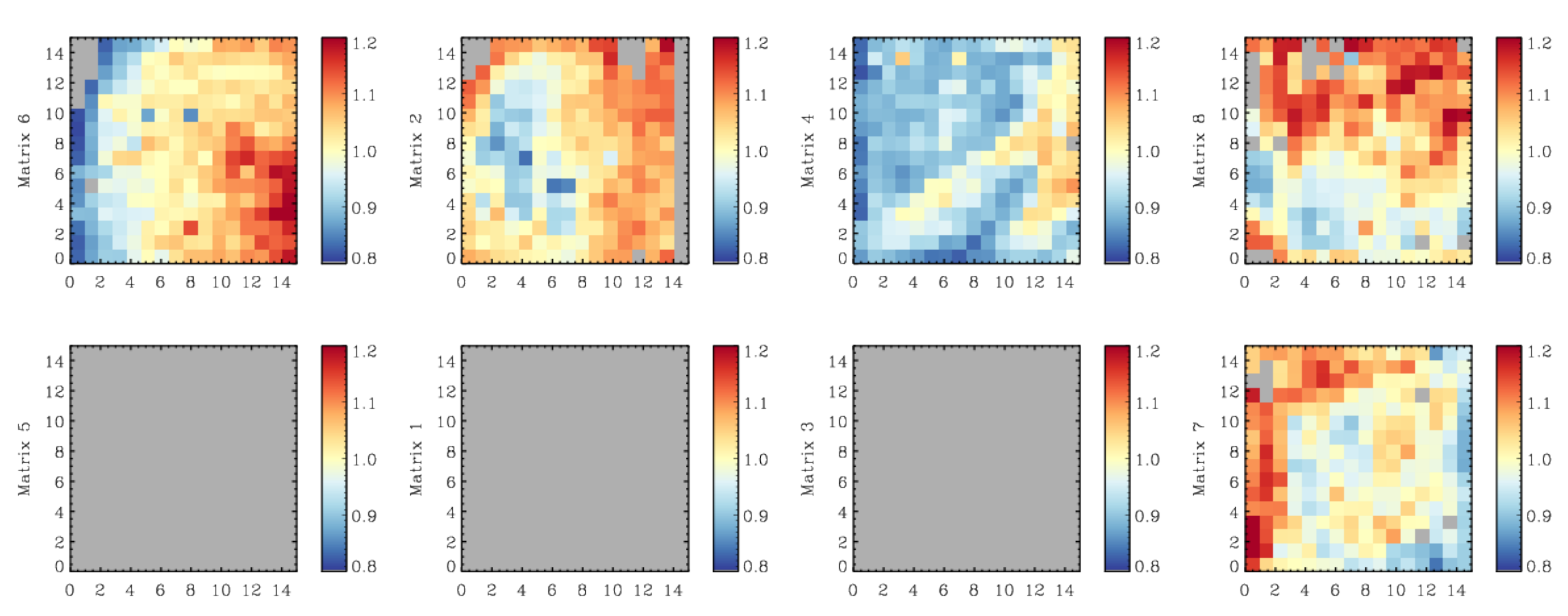}
\caption{Focal plane map of the mean response computed on the residual atmospheric signal during all scientific observations of {\flighttwo}. The response map has been divided by its average over valid detectors so that it has a mean equal to unity.}
\label{fig:response_map_scans}
\end{center}
\end{figure*}

The thermal emission from the residual atmosphere is clearly detected even at high altitude in the stratosphere. This strong signal is in principle extended and can therefore be used to measure the response flat-field of the detectors. In addition, since this signal is in principle unpolarized, it can be used to intercalibrate detectors of the {\TRANS} and {\REFLEX} focal planes.

We measure the detector response using the residual atmospheric emission as the slope of the correlation between the bolometer signal and the pointing elevation. This was done both for dedicated `skydip' measurements where the pointing elevation is changed continuously, and during normal science observations obtained at variable elevation angles during {\flighttwo}.  {The variations of the atmospheric emission signal with pointing elevation during {\flighttwo} was measured to be 8\,ADU/$\degr$ or $2\times10^{-3}$\,pW/pix/$\degr$ on average. These variations are thus be responsible for a signal increase of 0.18 pW from zenith to horizon. This figure can be compared to the absolute atmospheric emission towards zenith of 0.5 pW/pix, as estimated using the {\PILOT} photometric model \cite{Bernard_etal2016,Buttice2013}.}

Figures\,\ref{fig:response_map_skydips} and \ref{fig:response_map_scans} show the focal plane map of the mean response computed using the atmospheric signal during skydips and all observations of {\flighttwo}, respectively. Both maps present similar patterns. This pattern is also similar to that observed during ground test, when the focal planes were operated in front of an extended black-body, indicating that the structures observed are intrinsic pixel-to-pixel variations of the response across the focal planes. The accuracy of the response map is improved at the 1\% level or better when using all observations, due to the increased statistics with respect to skydip observations alone. This confirms the advantage of using a scanning strategy with varying elevation as implemented for {\flighttwo}.

\section{Detector noise}
\label{Sec:noise}

We present in this section the instrumental noise properties for {\flighttwo}. {We note that the noise properties during {\flightone} were similar to those of {\flighttwo} at night time, but were affected by significant external straylight caused by a deficit front baffle during the day. Note also that the high frequency noise level during {\flighttwo} was found to be significantly lower (up to 15\%) than during {\flightone}, which is likely due to operating the detectors at lower temperature (see Sect.\,\ref{sec:modif_between_flights}.}

\subsection{Flight-averaged noise power spectra}
\label{Sec:noise_averaged}

We compute the noise power spectra during {\flighttwo} for each detector and during each observation scan. For this purpose, the raw data are first corrected for the response time variations derived in Sect.\,\ref{Sec:ICS_resp} and converted to watts using the detector-averaged ground calibration factor of $2.116\,10^{10}$~V/W, derived from ground calibrations of the detectors alone in front of an absolute black body. The calibrated timelines are corrected for the atmospheric signal to first order, by removing their linear correlation with observation elevation as calculated over each observation, taking into account only scan data (i.e., excluding time samples associated with  calibrations, slews, etc.). Given the noise levels, we can assume that after the atmospheric signal removal, most of the individual detector timelines are dominated by instrumental noise. {We therefore did not discard the scans of our science targets and used all scans of {\flighttwo} in the present noise analysis.}

We compute the mean timeline among valid pixels of each array and scale it by the square root of the number of valid detectors in each array to keep a single-detector normalization. {We note that this procedure, while correctly characterizing the amplitude of the uncorrelated noise among detectors, might not correctly depict the relative amplitude of the correlated noise at the detector level, as the latter does not average out.} The array-averaged power spectra $P_{m,s}(\nu)$ are computed for each detector array $m$ and each scan $s$ from these array-averaged timelines, in ${\rm W}/\sqrt{\rm Hz}$, in the range of frequencies $\nu\in$[0.02,20]~Hz. Finally, we take the median value of $P_{m,s}(\nu)$ among the scans $s$ as the flight-averaged noise power spectra.

Similarly, we compute the half-pixel difference (HPD) noise power spectra by removing a common mode to all the detector timelines. For this purpose, we split each detector array in two subsets with the same number of pixels, chosen randomly, and compute the half-difference of the two detector-averaged timelines. From these HPD timelines, for each detector array and each scan, we compute the HPD power spectra $P_{m,s}^{\rm HPD}(\nu)$, which do not contain the common mode.

\begin{figure}[ht]
\begin{center}
\includegraphics[width=\columnwidth]{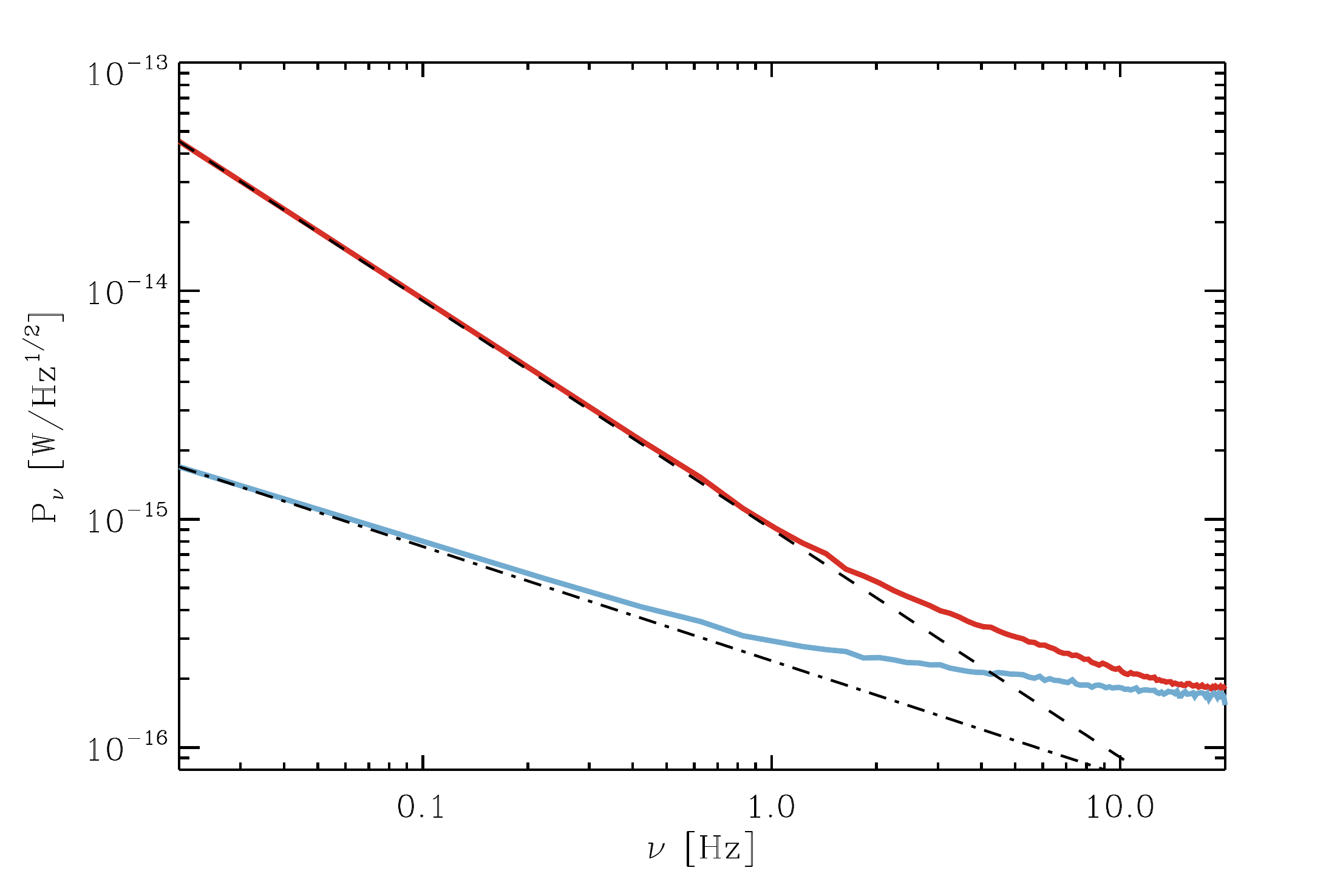}
\caption{\label{fig:flight-averaged-spectra} Array \#6 array-flight-averaged power spectrum $P_{6,s}(\nu)$ (red) and HPD flight-average $P_{6,s}^{\rm HPD}(\nu)$ (blue) in W/$\sqrt{\rm Hz}$, during {\flighttwo}. $1/\nu$ and $1/\sqrt\nu$ curves are overplotted to guide the eye as dashed- and dashed-dotted-lines, respectively.}
\end{center}
\end{figure}

The flight-averaged noise power spectra are shown in Fig.~\ref{fig:flight-averaged-spectra} for array \#6. The spectra for other detector arrays are qualitatively similar. For both the array-averaged spectrum $P_{m,s}(\nu)$ and the HPD spectrum $P_{m,s}^{\rm HPD}(\nu)$, two regimes can be identified. At high frequency, a flat spectrum component is observed, corresponding to white noise, presumably caused by photon noise {and uncorrelated between detectors}. At low frequency, a $1/\nu$ and a $1/\sqrt\nu$ component are observed for $P_{m,s}(\nu)$ and $P_{m,s}^{\rm HPD}(\nu)$, respectively. The former contains residuals from atmospheric emission and possibly variations of the focal plane temperature that contribute to the low-frequency rise of the spectrum{, that can be partially correlated between detectors}. In the HPD case, we can consider that all atmospheric and temperature variations are removed with the per-array common mode.

\subsection{Noise stability}

\begin{figure*}[ht]
\begin{center}
\includegraphics[width=\textwidth]{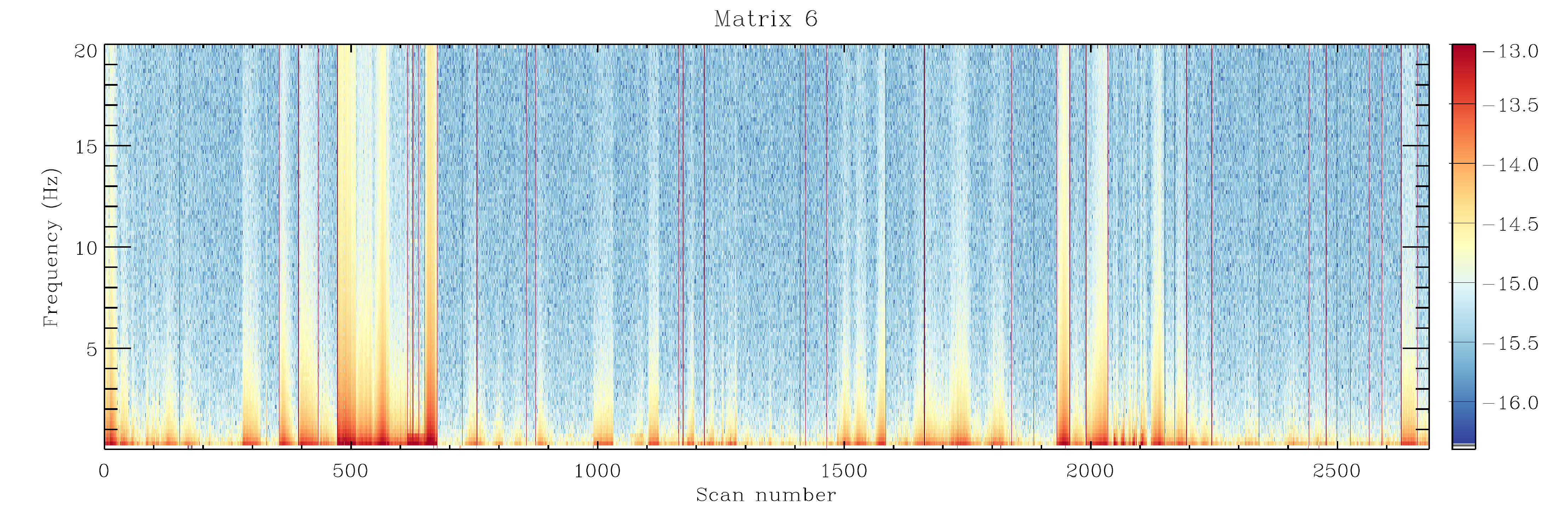}
\includegraphics[width=\textwidth]{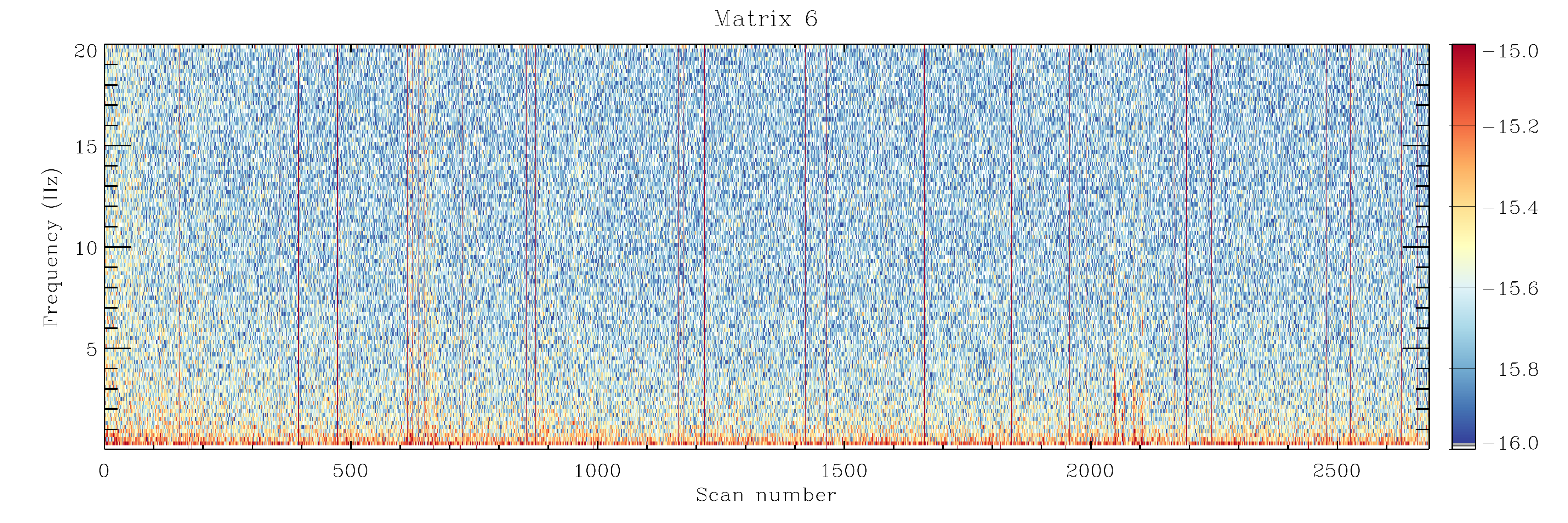}
\caption{\label{fig:t-nu_flight2} Time-frequency behaviour of the total array-averaged power spectra (top panel) and of the \emph{half-pixel difference} power spectra (HPD, bottom panel) for array \#6 during {\flighttwo}, in units of $\log_{10}({\rm W}/\sqrt{\rm Hz})$ {(note the different range in the top and bottom panels)}. Individual noise power spectra are computed for the array-average total signal or half-pixel differenced signal, for each individual observing scan.}
\end{center}
\end{figure*}

In order to study the noise stability during {\flighttwo}, we build time-frequency diagrams for $P_{m,s}(\nu)$ and $P_{m,s}^{\rm HPD}(\nu)$, shown for array \#6 in Fig.~\ref{fig:t-nu_flight2}. These diagrams are qualitatively similar for the other arrays.

In the array-averaged time-frequency diagram, some observations can be identified as having a larger low-frequency component. This is due to the simple atmospheric emission removal we have implemented here, which sometimes fails to properly subtract the low-frequency contribution. At higher frequency, a good stability of the white noise is observed (excluding these observations). In the HPD case, where the common mode for all the detectors belonging to the same array has been removed, the stability is remarkable for both low and high frequencies, during the whole duration of {\flighttwo}.

\subsection{High-frequency noise levels}

\begin{table*}
\begin{center}
{\footnotesize 
\begin{tabular}{llllll}
\hline\hline
ARRAY & MIN & MAX & AVG & MED & STDEV \\
& [$10^{-16}$ $\mathrm{W/\sqrt{Hz}}$] & [$10^{-15}$ $\mathrm{W/\sqrt{Hz}}$] & [$10^{-16}$ $\mathrm{W/\sqrt{Hz}}$] & 
[$10^{-16}$ $\mathrm{W/\sqrt{Hz}}$] & [$10^{-16}$ $\mathrm{W/\sqrt{Hz}}$]\\\hline
2              & 2.852 & 9.075 & 5.955 & 5.034 & 6.971\\ 
6              & 1.161 & 2.092 & 2.178 & 1.918 & 1.367\\\hline  
{\TRANS} average  & 2.006 & 5.584 & 4.066 & 3.476 & 4.169\\\hline 
7              & 3.419 & 8.083 & 6.598 & 6.287 & 4.994\\  
4              & 2.722 & 2.111 & 4.665 & 4.586 & 1.354\\  
8              & 3.144 & 6.936 & 5.578 & 5.196 & 4.364\\\hline 
{\REFLEX} average & 3.095 & 5.710 & 5.614 & 5.356 & 3.571\\\hline  
Average        & 2.659 & 5.659 & 4.995 & 4.604 & 3.810\\\hline
\end{tabular}
}
\end{center}
\caption{\label{tab:wn_flight} High frequency noise statistics for each array during {\flighttwo} $\mathrm{W/\sqrt{Hz}}$.}
\end{table*}

To assess the high-frequency noise level statistics during {\flighttwo}, we repeat the initial steps presented in Sect.~\ref{Sec:noise_averaged}. However, instead of averaging the signal among the detectors for each array, we compute the noise power spectra $P_{i,s}(\nu)$ for each detector $i$ and each scan $s$. For each detector, we compute the flight-averaged spectrum $P_{i}(\nu)$ as the median over all the scans. The high frequency noise levels are then taken to be the mean value of $P_{i}(\nu)$ in the range $\nu\in$[0.02,20]~Hz.

The statistics of these high-frequency noise levels are presented in Tab.~\ref{tab:wn_flight}. The focal plane median high-frequency noise level is $4.6\,10^{-16}$~W$/\sqrt{\rm Hz}$. Array \#6 is the most sensitive with a median sensitivity corresponding to a high-frequency noise level of $1.9\,10^{-16}$~W$/\sqrt{\rm Hz}$. Arrays \#2, \#4, \#7 and \#8 have similar sensitivities (ranging from 4.5 to 6$\,10^{-16}$ W$/\sqrt{\rm Hz}$).

\section{Expected sensitivities}
\label{Sec:sensitivity}

\begin{figure}[ht]
\begin{center}
\includegraphics[width=1.0\columnwidth,angle=0]{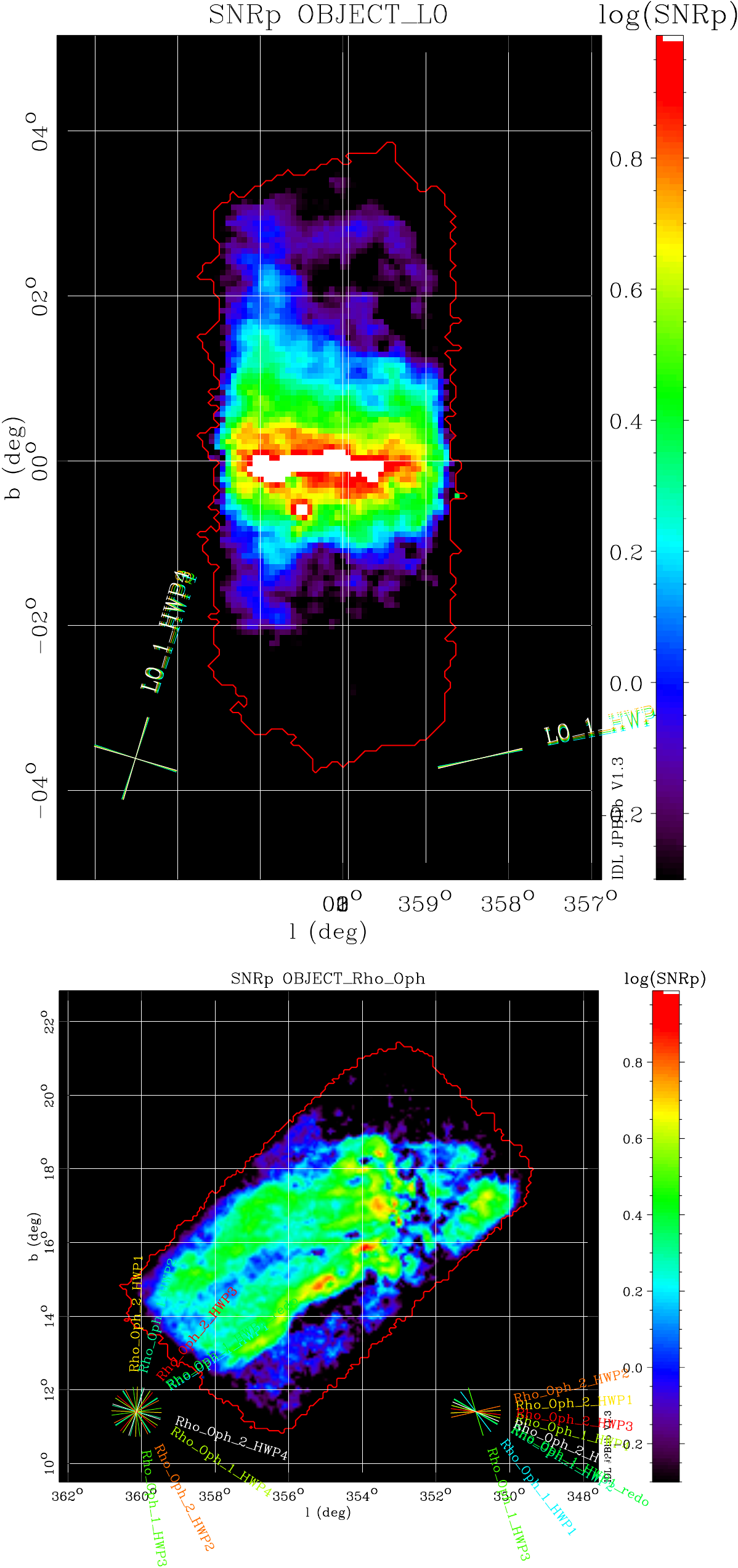}
\end{center}
\caption{\label{fig:SNRp} Maps of the expected SNR ratio on the polarization fraction $p$ for the Galactic center region at 2$^\prime$ resolution (top) and the Rho-Ophiuchi region (bottom) at 5$^\prime$ resolution.}
\end{figure}

We present the signal to noise ratio predictions for the polarization fraction $p$ ($\SNRp$), based on the simulations that are done with the {\flighttwo} observing strategy and parameters. The expected sensitivities are calculated as described in \cite{Bernard_etal2016}, based on simulations where the Stokes parameters $\StokesI$, $\StokesQ$, $\StokesU$ are taken from the {\Planck} polarization maps at 353 GHz \cite{planck2014-XIX}, and extrapolated to the {\PILOT} frequency using a modified black body spectrum with a dust emissivity index and dust temperatures also determined from analysis of the {\Planck} data (\cite{planck2013-XI}){, and also assuming the polarization fraction measured with {\Planck}}. We use the median value of the high frequency noise as measured during {\flighttwo}, corresponding to a total NEP$= 4.6\,10^{-16}$~W$/\sqrt{\rm Hz}$ (see Tab. \ref{tab:wn_flight}). These predictions assume that the accuracy of the final maps will be limited by the high frequency noise, as expected if the systematic effects are fully suppressed and the map-making is optimal. Figure \ref{fig:SNRp} shows the maps of the expected SNRp for the Galactic Center and the Rho-Ophiuchi regions. We expect to obtain $\SNRp\simeq 10$ on the weakly polarized region of the Galactic Center at 2$^\prime$ resolution and a $\SNRp\simeq 6$ for the bright Rho-Oph sources at 5$^\prime$ resolution. We expect $SNRp\simeq 16$ when integrating over the whole {\BICEPtwo} diffuse field, when assuming 20\% of polarization in agreement with the Planck measurement at 353GHz. The observations of the Large Magellanic Cloud will allow us to study the polarisation in this region at intermediate scales (30$^\prime$).

\section{Conclusions}
\label{sec:conclusion}

The {\Pilot} experiment has had two successful flights, one from Timmins, Ontario Canada in September 2015, and from Alice Springs, Australia in April 2017.  The science observations during these flights targeted nearby star-forming regions, molecular clouds, cold cores, the Local Group galaxies M31 and LMC, and diffuse ISM regions, including the BICEP2 region. We  also observed planets and performed skydips for calibration purposes.  Our analysis of the house-keeping and scientific data indicates nominal behavior of the instrument. We accurately measured the time constant of the {\PILOT} detectors using glitches and the internal calibration source. The optical quality was checked on planets and found to be consistent with expectations. The FWHM of the {\PILOT} PSF is $2^\prime.25$. The intensity of the instrumental background during flight is difficult to measure accurately, but we conclude that it is likely higher than the prediction by our photometric model. Our measurements from ground calibration tests and during flight indicate that instrumental background is polarized. The background polarization fraction in the flight data is at a similar level as in the data from ground calibration tests, but the orientation of the instrumental polarization between the flight and ground data differs {by 35$\degr$}. The polarization of the instrumental background in combination with residual atmosphere emission mean that the instrumental background varies during flight. The response of the detectors, as measured in-flight using the {\ICS} signal, is found to be variable, which we ascribe to variations of the background on the detectors and to variations in the focal plane temperature. A simple parametric model based on house-keeping measurements reproduces the observed variations of the response to a few percent. {The inflight detector noise is at the expected level, with low frequency noise rising as $1/\sqrt\nu$ when a common mode is removed, and a flattening observed at high frequency.}  The spectral shape and amplitude of the noise are found to be stable during flight. Having obtained a quantitative characterization of the in-flight performance of {\PILOT}, we re-evaluated its expected sensitivity to polarization, which we find is sufficient to fulfil our science goals.

\begin{acknowledgements}
{\PILOT} is an international project that involves several European institutes.  The institutes that have contributed to hardware developments for {\PILOT} are IRAP and CNES in Toulouse (France), IAS in Orsay (France), CEA in Saclay (France), Rome University in Rome (Italy) and Cardiff University (UK).  This work was supported by the CNES. It is based on the {\Pilot} data obtained during two flight campaigns operated by CNES, under the agreement between CNES and CNRS/INSU. This work was supported by the Programme National ``Physique et Chimie du Milieu Interstellaire'' (PCMI) of CNRS/INSU with INC/INP co-funded by CEA and CNES. We thank S. Hanany for his support during the Australian flight campaign and useful discussions about the results presented here. 

\end{acknowledgements}

\bibliographystyle{spphys}       
\bibliography{perfo}   

\end{document}